\documentclass[10pt,a4paper,notitlepage]{article}
\usepackage[utf8]{inputenc}
\usepackage{amsmath,amsfonts,amssymb}
\usepackage{graphicx}
\usepackage{hyperref}
	\usepackage[english]{babel}
\usepackage{epstopdf}
\usepackage{multirow}
\usepackage{tabularx}
\usepackage{floatrow}
\usepackage{subcaption}

\hypersetup{
	colorlinks=true,         
	linkcolor=blue,          
	citecolor=Violet,        
	urlcolor=Violet            
}
\usepackage[usenames,dvipsnames]{color}

\begin{document}

	\title{ 	
		\Large \bf{Non-Abelian axial anomaly, axial-vector duality, \\
			and the pseudoscalar glueball} }
	
	\author{Sergey~Khlebtsov,$^1$~~
		Yaroslav~Klopot,$^{2,3}$ ~~ 
		Armen~Oganesian,$^{1,2}$~~
		Oleg~Teryaev,$^2$
		\footnote{Electronic addresses: 
			\href{mailto:khlebtsov@itep.ru,}{khlebtsov@itep.ru},
			\href{mailto:klopot@theor.jinr.ru}{klopot@theor.jinr.ru}, \href{mailto:armen@itep.ru,}{armen@itep.ru,} \href{mailto:teryaev@theor.jinr.ru}{teryaev@theor.jinr.ru}.}
		\vspace{12pt} \\
		\it \small  $^1$Institute of Theoretical and Experimental Physics, 117218,  Moscow, Russia\\
		\it \small $^2$ 
		\it \small Joint Institute for Nuclear Research, 141980, Dubna, Russia\\
		\it \small $^3$Bogolyubov Institute for Theoretical Physics, 03143, Kyiv, Ukraine}
	
	\date{}
	\maketitle

	\begin{abstract}
		We generalize the dispersive approach to axial anomaly by A.D. Dolgov and V.I. Zakharov \cite{Dolgov:1971ri} to a non-Abelian case with arbitrary photon virtualites. We derive the anomaly sum rule for the singlet current and obtain the $\pi^0,\eta,\eta'\rightarrow\gamma\gamma^*$ transition form factors. Using them, we established the behavior of a nonperturbative gluon matrix element $\langle 0 |G\tilde{G} |\gamma\gamma^*(q^2) \rangle$ in both spacelike and timelike regions. We found a significant contribution of the non-Abelian axial anomaly to the processes with one virtual photon, comparable to that of the electromagnetic anomaly. The duality between the axial and the vector channels was observed: the values of duality intervals and mixing parameters in the axial channel were related to vector resonances' masses and residues. The possibility of a light pseudoscalar glueball-like state is conjectured. 
	\end{abstract}
	\section{ Introduction} 
	
	One of the fundamental features of QCD, the axial anomaly, has many theoretical and phenomenological applications. 
	Being essentially a nonperturbative phenomenon of QCD, the axial anomaly provides valuable insights into low-energy properties of the theory, inaccessible to perturbative methods.
	Among phenomenological manifestations of special importance are the $\gamma\gamma$ decays of neutral pseudoscalar mesons $\pi^0, \eta, \eta'$. It was the problem of the neutral pion decay, which, in fact, led to the discovery of quantum anomalies \cite{Adler:1969gk,Bell:1969ts}. The processes involving virtual photons, like transition form factors (TFFs) of pseudoscalar mesons, are also deeply connected with the axial anomaly. 
	
	The existence of the axial anomaly results in the nonconservation of the axial current. In particular, a singlet axial current  $J_{\mu5}^{(0)}=(1/\sqrt{3})\sum_i {\bar{\psi_i}\gamma_{\mu}\gamma_5 \psi_i}$ gets both electromagnetic and strong (non-Abelian) anomalies, while isovector ($a=3$) and octet ($a=8$) components of the octet of axial currents  $J_{\mu5}^{(a)}=(1/\sqrt{2})\sum_i {\bar{\psi_i}\gamma_{\mu}\gamma_5\lambda^a \psi_i}$  acquire only the electromagnetic anomaly:
	
	\begin{equation}\label{an-0}
	\partial^\mu J_{\mu 5}^{(0)} =\frac{2i}{\sqrt 3}\sum_i{m_i \bar{\psi_i} \gamma_5 \psi_i}+\frac{e^2}{8\pi^2}C^{(0)}N_c  F\widetilde{F} + \frac{\sqrt{3}\alpha_s}{4\pi}  G\widetilde{G},
	\end{equation}
	
	\begin{equation}\label{an-38}
	\partial^\mu J_{\mu 5}^{(a)} =\frac{2i}{\sqrt2}\sum_i{m_i \bar{\psi_i} \gamma_5 \lambda^a \psi_i} 
	+\frac{e^2}{8\pi^2}C^{(a)}N_c  F\widetilde{F}, \; a=3,8.
	\end{equation}
	Here $F$ and $G$ are electromagnetic and gluon field strength tensors respectively,  $\widetilde{F}^{\mu\nu}=\frac{1}{2}\epsilon^{\mu\nu\rho\sigma}F_{\rho\sigma}$ and $\widetilde{G}^{\mu\nu,t}=\frac{1}{2}\epsilon^{\mu\nu\rho\sigma}G_{\rho\sigma}^{t}$ are their duals, $N_c=3$ is a number of colors, $\alpha_s$ is a strong coupling constant, $C^{(a)}$ are the charge factors ($e_i$ are quark charges in units of the electron charge $e$): $C^{(3)}=\frac{1}{\sqrt 2} (e_u^2-e_d^2)=\frac{1}{3\sqrt 2}, C^{(8)}=\frac{1}{\sqrt 6} (e_u^2+e_d^2-2e_s^2)=\frac{1}{3\sqrt 6}$, $C^{(0)}=\frac{1}{\sqrt 3} (e_u^2+e_d^2+e_s^2)=\frac{2}{3\sqrt 3}.$  The sum is over the flavors of light quarks $i=u,d,s$; $\lambda^a$ are the diagonal Gell-Mann $SU(3)$ matrices, $a=3,8$.

	The axial anomaly can be calculated as a result of ultraviolet regularization of the diverging triangle VVA diagram. Alternatively, it can be derived considering the imaginary part of the VVA diagram \cite{Dolgov:1971ri}, where the anomaly arises as a sum rule for a structure function in the dispersion representation of the three point VVA correlation function \cite{Horejsi:1985qu,Horejsi:1994aj,Veretin:1994dn} (for a review, see \cite{Horejsi:1992tx,Ioffe:2006ww}). Such sum rules, combined with the global quark-hadron duality, then can be employed to study the $\gamma\gamma$ decays of the pseudoscalar mesons as well as their transition form factors $\gamma\gamma^* \rightarrow \pi^0,\eta,\eta'$. In particular, the anomaly sum rule (ASR) for the isovector axial current was related to the pion TFF \cite{Klopot:2010ke}, while the anomaly sum rule for the octet axial current was developed and used to study the $\eta, \eta'$ TFFs, including the cases of spacelike \cite{Klopot:2011qq, Klopot:2011ai, Melikhov:2012qp, Klopot:2012hd} and timelike \cite{Klopot:2013laa} photon virtualities. These anomaly sum rules for the isovector ($a=3$) and octet ($a=8$) currents when one of the photons is real ($k^2=0$) and another is real or virtual ($Q^2=-q^2 \geq 0$) read (in what follows we put $m_u=m_d=m_s=0$) \cite{Horejsi:1994aj,Klopot:2010ke,Klopot:2011qq},
	
	\begin{equation}\label{asr38}
	\frac{1}{\pi}\int_{0}^{\infty}A^{(3,8)}_3(s,Q^2)ds = \frac{C^{(3,8)}N_c}{2\pi^2},
	\end{equation}
	where the spectral density function is defined as $A^{(3,8)}_{3} \equiv \frac{1}{2}Im(F_3-F_6)$ and is determined from a decomposition of the vector-vector-axial (VVA) amplitude
	
	\begin{equation} \label{VVA}
	T_{\alpha \mu\nu}(k,q)=e^2\int
	d^4 x d^4 y e^{(ikx+iqy)} \langle 0|T\{ J_{\alpha 5}(0) J_\mu (x)
	J_\nu(y) \}|0\rangle
	\end{equation}
	as \cite{Rosenberg:1962pp} (see also \cite{Eletsky:1982py,Radyushkin:1996tb}) 
	\begin{align}
	\label{eq1} \nonumber T_{\alpha \mu \nu} (k,q)  & =  F_{1} \;
	\varepsilon_{\alpha \mu \nu \rho} k^{\rho} + F_{2} \;
	\varepsilon_{\alpha \mu \nu \rho} q^{\rho} + F_{3} \; k_{\nu} \varepsilon_{\alpha \mu \rho \sigma}k^{\rho} q^{\sigma}
	\\ 
	& + F_{4} \; q_{\nu} \varepsilon_{\alpha \mu \rho \sigma} k^{\rho}q^{\sigma} + F_{5} \; k_{\mu} \varepsilon_{\alpha \nu \rho \sigma} k^{\rho} q^{\sigma} + F_{6} \; q_{\mu} \varepsilon_{\alpha \nu \rho \sigma} k^{\rho} q^{\sigma},
	\end{align}
	where the coefficients $F_{j} = F_{j}(p^{2},q^{2})$, $j = 1, \dots ,6$ are the corresponding Lorentz invariant amplitudes constrained by current conservation and Bose symmetry. The electromagnetic currents are defined as $J_{\mu}=\sum_i {e_i\bar{\psi_i}\gamma_{\mu}\psi_i}$ with momenta $k,q$.
	The rhs of (\ref{asr38}) is exactly the Abelian (electromagnetic) anomaly constant stemmed from the matrix element $\langle 0 |F\tilde{F} |\gamma\gamma^* \rangle$.
	
	Due to appearance of the strong anomaly term in the singlet current (\ref{an-0}), the analogous anomaly sum rule has an additional part stemmed from the matrix element 
	\begin{equation} \label{N}
	\langle 0 | \frac{\sqrt{3}\alpha_s}{4\pi} G\tilde{G}|\gamma\gamma \rangle = N(p^2, k^2,q^2) \epsilon^{\mu\nu\rho\sigma}k_{\mu}q_{\nu}\epsilon_{\rho}^{(k)}\epsilon_{\sigma}^{(q)}.
	\end{equation}
	The corresponding non-Abelian contribution in the dispersive form requires a subtraction, so the singlet current ASR in the considered kinematics $N(p^2, k^2=0, q^2=-Q^2)\equiv N(p^2, Q^2)$ can be written  \cite{Khlebtsov:2018roy} as  
	
\begin{equation}\label{asr0}
\frac{1}{\pi}\int_{0}^{\infty}A^{(0)}_{3}(s,Q^{2})ds = \frac{C^{(0)}N_c}{2\pi^2} + N(0,Q^2)-\frac{1}{\pi}\int_{0}^{\infty}Im R(s,Q^2)ds,
\end{equation}
where 
\begin{equation}\label{N_sub}
	R(p^2,Q^2)=\frac{1}{p^2}(N(p^2,Q^2)-N(0,Q^2)).
\end{equation}

Therefore, in addition to the Abelian anomaly contribution, the rhs of Eq. (\ref{asr0}) has also a non-Abelian anomaly contribution (\ref{N}) given by the subtraction $N(0, Q^2)$ and the spectral parts. Rigorous calculation of these form factors face difficulty due to their nonperturbative nature. However, they can be related to the physical observables by means of the quark-hadron duality hypothesis. To carry this out, we saturate the lhs of ASRs (\ref{asr38}) and (\ref{asr0}) with a full set of resonances and single out the lowest-lying contributing states in each channel in terms of the corresponding transition form factors and decay constants. The ``continuum" contribution absorbs the rest (higher resonances)  with the same function $A_3(s,Q^2)$.  So, we apply the following quark-hadron model,

\begin{equation}\label{qhd}
A^{(a)}_3(s,Q^2)=\pi  \Sigma f_P^{(a)}(s)\delta(s-m^{2}_P) F_{P\gamma}(s, Q^2)+A_3^{(a)}(s,Q^2) \theta(s-s_{a}),  \; a=3, 8, 0.
\end{equation}
Here the sum is over the hadron states $P$ whose decay constants $f_P^{(a)}$ and the form factors $F_{P\gamma}$ of the transitions $\gamma\gamma^* \to P$ are defined as
\begin{equation} \label{def_f}
\langle 0|J^{(a)}_{\alpha 5}(0) |P(p)\rangle=
i p_\alpha f^{(a)}_P(p^2), 
\end{equation}
\begin{equation} \label{def_tff}
\int d^{4}x e^{ikx} \langle P(p)|T\{J_\mu (x) J_\nu(0)
\}|0\rangle = e^2\epsilon_{\mu\nu\rho\sigma}k^\rho q^\sigma
F_{P\gamma}(p^2,Q^2)\;. 
\end{equation}

Then the ASRs  (\ref{asr38}) and (\ref{asr0}) read,
\begin{equation}\label{asr38res}
\Sigma f_P^{(3,8)}(m_P^2)F_{P\gamma}(m_P^2,Q^2)+\frac{1}{\pi}\int_{s_{3,8}}^{\infty}A^{(3,8)}_3(s,Q^2)ds = \frac{C^{(3,8)}N_c}{2\pi^2},
\end{equation}
\begin{equation}\label{asr0res}
\Sigma f_P^{(0)}(m_P^2)F_{P\gamma}(m_P^2,Q^2)+\frac{1}{\pi}\int_{s_0}^{\infty}A_3^{(0)}ds = \frac{C^{(0)}N_c}{2\pi^2} + N(0,Q^2)-\frac{1}{\pi}\int_{0}^{\infty}Im R(s,Q^2)ds.
\end{equation}

The lowest hadron contributions to the ASRs \eqref{asr38res} and \eqref{asr0res} are given by the $\pi^0, \eta$ and $\eta'$ mesons, while the rest of the hadrons are absorbed by the integrals with lower limits $s_3$, $s_8$, $s_0$ -- the duality intervals in the respective channels. The numerical values of these parameters will be discussed in the next section. 
	
	It is important to note that the singlet current is renormalization-scale dependent (unlike the isovector and octet currents) resulting in $\sim$ 15$\%$ decrease of $f_P^{(0)}$ at high $Q^2$, which was recently taken into account in analyses of the TFFs \cite{Agaev:2014wna,Escribano:2015nra,Escribano:2015yup}.  Nevertheless,  in the ASR \eqref{asr0res},  the decay constants are fixed on the respective meson mass scales as a result of  the $\delta$-like resonance model in (\ref{qhd}). 
	
	\begin{figure}[H]\label{fig:1}
		\ffigbox{
			\begin{subfloatrow}[3]
				\ffigbox[\FBwidth]{\caption{} \label{fig:2}}%
				{\includegraphics[width=0.33\textwidth]{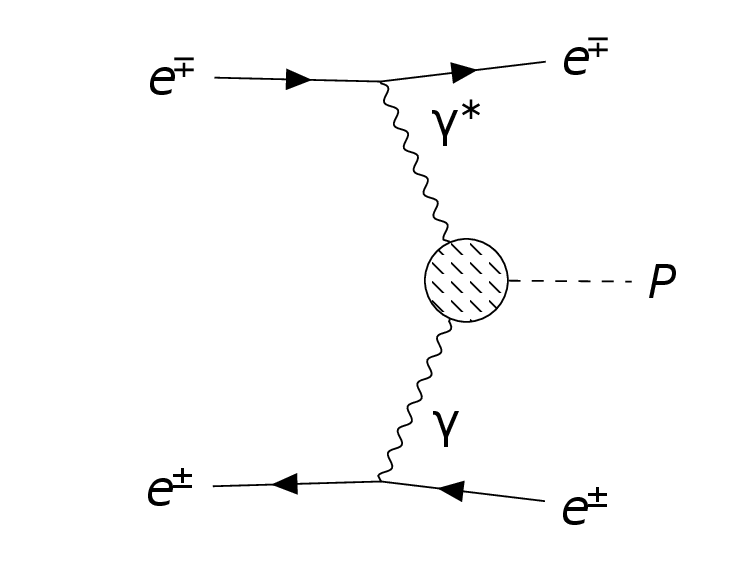}}
				\ffigbox[\FBwidth]{\caption{}\label{fig:3}}%
				{\includegraphics[width=0.33\textwidth]{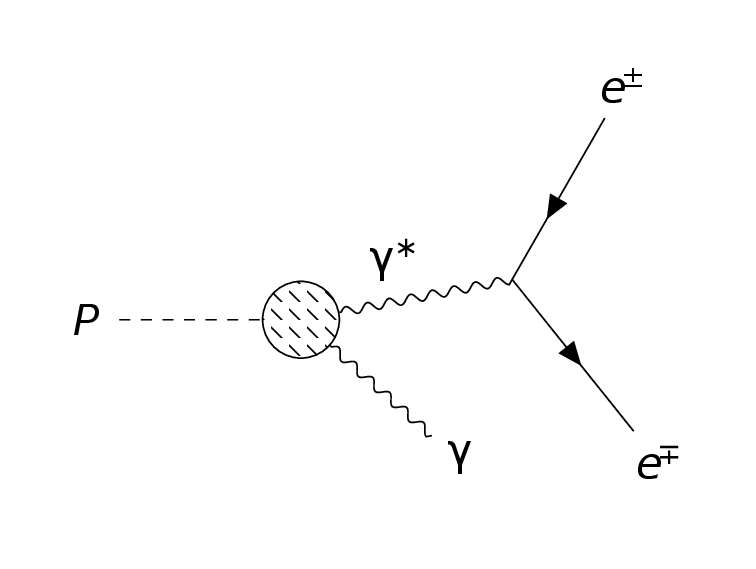}}  
				\ffigbox[\FBwidth]{\caption{} \label{fig:4}}%
				{\includegraphics[width=0.33\textwidth]{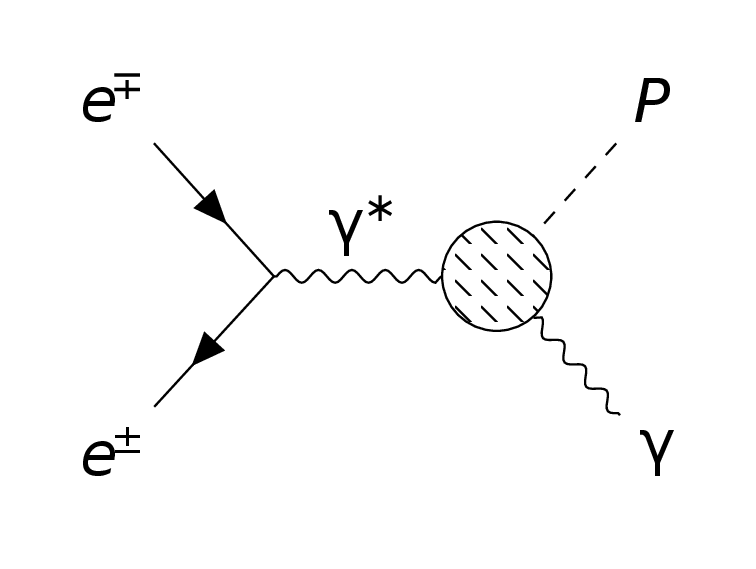}}
			\end{subfloatrow}
		}
		{
			\caption{The Feynman diagrams of the processes: (a)  $e^{+}e^{-}\rightarrow e^{+}e^{-}P$ scattering, (b) $P\rightarrow e^{+}e^{-}\gamma$ Dalitz decays, (c)  $e^{+}e^{-}\rightarrow P\gamma$ annihilation.}
		}
	\end{figure}
	
	Regarding the experiment, the transition form factors  $P\rightarrow\gamma^*\gamma$ ($P = \pi^0,\eta,\eta'$) are well studied. The spacelike region of virtual photon is studied in the two-photon fusion process $e^+e^-\rightarrow e^+e^-P$ (Fig. \ref{fig:2}). A large amount of data covering the region of $Q^2=-q^2>0$ from 0.06 up to 40 GeV$^2$ was obtained by the collaborations L3 \cite{Acciarri:1998}, CELLO \cite{Behrend:1990sr}, CLEO \cite{Gronberg:1997fj}, BaBar \cite{Aubert:2009mc,BABAR:2011ad}, and BELLE \cite{Uehara:2012ag}. 
	The timelike region was investigated via Dalitz decays ($4m_l^2\leq q^2<m^2_{P}$, Fig. \ref{fig:3}) and in the annihilation of $e^+e^-$ to hadrons ($q^2\geq m^2_{P}$, Fig. \ref{fig:4}). The Dalitz decays were measured by A2 \cite{Adlarson:2016ykr,Adlarson:2016hpp}, NA60 \cite{Arnaldi:2016pzu} and BESIII \cite{Ablikim:2015wnx} collaborations. The cross sections of $e^{+}e^{-}\rightarrow P\gamma$ were measured by SND \cite{Achasov:2016bfr,Achasov:2007kw, Achasov:2013eli,Achasov:2018ujw}, CMD \cite{Akhmetshin:2004gw} and BaBar \cite{Aubert:2006cy} collaborations covering the region of  $\sqrt{q^2}=\sqrt{s}$ from 0.017 up to 2 GeV. The BaBar collaboration measured also the cross sections at $q^2=112$  GeV$^2$  \cite{Aubert:2006cy}. 
	In the future, the new data are expected from the BELLE, BESIII, CMD, and SND collaborations. Also KLOE2 \cite{Babusci:2011bg} and CLAS \cite{Amaryan:2013eja} reported their plans to study the $\pi^0,\eta,\eta'$ mesons.

	The TFFs of pseudosaclar mesons have also been a subject of extensive theoretical study within various frameworks, such as light cone sum rules \cite{Agaev:2014wna,Stefanis:2012yw, Mikhailov:2016klg},  constituent \cite{Dorokhov:2013xpa}, light-front \cite{Choi:2017zxn} and nonlocal chiral quark models \cite{GomezDumm:2016bxp}, light-front holographic QCD \cite{Brodsky:2011xx} as well as some other approaches \cite{Escribano:2015yup,Roig:2014uja,Nedelko:2016vpj,Czyz:2017veo,Escribano:2013kba,Hanhart:2013vba}. 
	
	Recently, the ASR for the singlet current was used to study the role of strong and electromagnetic anomalies in the processes with real photons $\eta,\eta\rightarrow \gamma\gamma$ \cite{Khlebtsov:2018roy}. We found that the strong anomaly contribution to these processes is small in comparison with the electromagnetic one.
	
	In this paper, using the same framework, we are going to generalize the obtained results and consider the case of one virtual photon. Combining ASRs for all three currents and incorporating the quark-hadron duality, we derive relations between the pseudoscalar TFFs and matrix element  $\langle 0 |G\tilde{G} |\gamma\gamma^* \rangle$. Then, using the experimental data for the TFFs, we study the properties of this matrix element at different photon virtualities.
	
	The paper is organized as follows: in Sec. \ref{sec2}, the hadrons contributions and duality intervals are studied.  Section \ref{sec3} is dedicated to analysis of the strong anomaly contribution in the spacelike region. The analytic continuation of the ASRs to the timelike region is considered and the investigation of the strong anomaly properties in this region is carried out in Sec. \ref{sec4}. The last Sec. \ref{last}  summarizes the results.
	
	\section{Hadron contributions and duality intervals}\label{sec2}
	
	The lower integration limits $s_3,s_8,s_0$ in \eqref{asr38res}, \eqref{asr0res}, emerging as free parameters of the ASR approach, are in fact the duality intervals (continuum thresholds) of the corresponding currents.
	
	In the isovector and octet currents, the duality intervals $s_3$, $s_8$ can be determined from the large $Q^2$ limit of the ASRs (\ref{asr38}) by making use of the pQCD asymptotes of the $\pi_0$, $\eta,\eta'$ TFFs  and numerical values of the decay constants $f_P^{(3,8)}$ \cite{Klopot:2011qq}. In this way, the interval of duality of the isovector channel is evaluated as $s_3\simeq 0.67$ GeV$^2$. 
	However, this parameter can have a weak $Q^2$ dependence \cite{Oganesian:2015ucv,Khlebtsov:2016vyf} varying from $0.6$ GeV$^2$ at $Q^2\to 0$ to $0.67$ GeV$^2$ at $Q^2 \to \infty$.
	In the case of the octet channel, the numerical value of the corresponding interval of duality was found to vary $s_8=0.4-0.6$ GeV$^2$, depending on a particular mixing scheme  \cite{Klopot:2011ai, Klopot:2012hd}. 
	However, in the chiral limit ($m_q=0$) it is natural to assume that the duality intervals of the isovector and octet currents cannot be much different from each other: $s_8\simeq s_3$ within $20\%$ uncertainty of the $SU(3)$  symmetry breaking.  For the purposes of numerical analysis, we take $s_8\simeq s_3 =0.6$ GeV$^2$ in this paper.
	
	The duality interval of the singlet current $s_0$ is different from $s_3$ and $s_8$. The main hadron contribution here is given by the $\eta'$ meson, which is known to retain its mass in the chiral limit, contrary to the lighter $\pi^0$ and $\eta$ mesons. So, the value of $s_0$ should be of order of $m^2_{\eta'}$ and it is natural to assume that $s_0 \simeq 1$ GeV$^2$. Although it is possible that the value of $s_0$ can have a weak $Q^2$ dependence, similar to that of $s_3$.
	
\begin{figure}[H]
		\begin{floatrow}
			\ffigbox{\caption{$A_{QED}$.} \label{fig:5}}%
			{\includegraphics[width=0.5\textwidth]{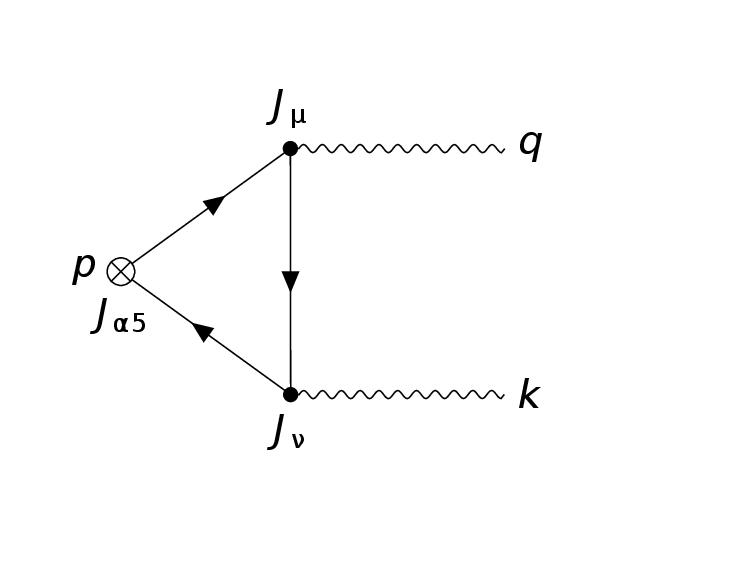}}
			\ffigbox{\caption{$A_{QCD}$.}\label{fig:6}}%
			{\includegraphics[width=0.5\textwidth]{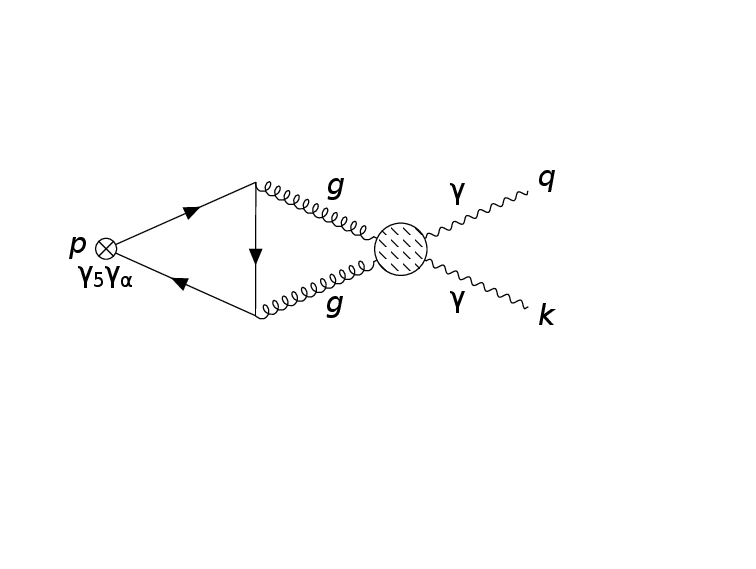}}  \end{floatrow}
	\end{figure}

	The one-loop approximation for the spectral densities of the isovector and octet currents $A_3^{(3,8)}(s,Q^2)$ (Fig. \ref{fig:5})  in the chiral limit is given by \cite{Radyushkin:1996tb} 
	
	\begin{align} \label{a3}
	A_{3}^{(3,8)}(s, Q^2)=\frac{C^{(3,8)}N_c}{2\pi}\frac{Q^2}{(s+Q^2)^2},
	\end{align}
	so the integration in the ASRs (\ref{asr38res}) leads to the following expressions for the hadron contributions,
	\begin{equation} \label{asr38aa}
	\Sigma f_P^{(3,8)}F_{P\gamma}(Q^2) = \frac{C^{(3,8)}N_c}{2\pi^2}\frac{s_{3,8}}{s_{3,8}+Q^2}.
	\end{equation}

	The case of the singlet current differs from the isovector and octet currents due to a new type of diagrams involving virtual gluons. 
	In order to single out electromagnetic contribution, we split the spectral density into two parts,
	\begin{equation}
	A_3^{(0)} = A_{QED}^{(0)}+A_{QCD}^{(0)}.
	\end{equation}
	The second part $A_{QCD}^{(0)}$ is the contribution of diagrams (Fig.~\ref{fig:6}) with virtual gluons coupled to two photons through nonperturbative strong interactions (see also \cite{Khlebtsov:2018roy}). The first part $A_{QED}^{(0)}$ represents the contribution of QED diagrams, whose lowest one-loop part (Fig. \ref{fig:5}) is given by a similar expression to Eq. (\ref{a3}) with an appropriate charge factor $C^{(0)}$. Making use of it, we can rewrite the ASR \eqref{asr0res} as
	\begin{equation}\label{qedplusqcd}
	\Sigma f_P^{(0)}F_{P\gamma}(Q^2) = \frac{N_cC^{(0)}}{2\pi^2}\frac{s_0}{s_0+Q^2} - \frac{1}{\pi}\int_{s_0}^{\infty} A_{QCD}ds + N(0,Q^2) - \frac{1}{\pi}\int_{0}^{\infty} ImR(s,Q^{2})ds.
	\end{equation} 
	The first and the last three terms in Eq. (\ref{qedplusqcd}) represent the electromagnetic and the strong anomaly contributions to the ASR respectively. It is convenient to introduce a function that represents the ratio of contributions of strong and electromagnetic anomalies:
	\begin{equation}\label{gluon_anom_contrib1}
	B(Q^2, s_0) = \frac{2\pi^2}{N_cC^{(0)}}\frac{s_0+Q^2}{s_0} \left[ N(0,Q^2) - \frac{1}{\pi}\int_{0}^{\infty} ImR(s,Q^{2})ds - \frac{1}{\pi}\int_{s_0}^{\infty} A_{QCD}(s,Q^2)ds \right].
	\end{equation} 
	
	As the integral of $A_{QCD}$ is suppressed as $\alpha^2_s$ at $s_0 \geq 1.0$ GeV$^2$, the function $B(Q^2,s_0)$ is predominantly determined by the first two terms. It reflects the properties of the nonperturbative matrix element  $\langle 0 |G\tilde{G}|\gamma\gamma^* \rangle$. \footnote{
		Let us note, that in our approach, based on quark-hadron duality, the EM anomaly contribution is parametrically dependent on the value of $s_0$ and so the function $B(Q^2,s_0)$ has also dependence on $s_0$.}
Therefore, the study of the function $B(Q^2,s_0)$ gives us access to the non-Abelian anomaly contribution to the processes $P\to \gamma^*\gamma$ at various photon virtualities, which is the main goal of the present paper.
	
	As $B(Q^2,s_0)$ cannot be calculated perturbatively, we will extract it from the experimental data. This resembles the extraction of nonperturbative ingredients, like (transverse momentum dependent) parton-distribution functions in QCD and (nonlocal) vacuum condensates from data or lattice simulations \cite{Kotov:2019dby}. In a sense, we are going to solve the inverse problem.

	So, rewriting the ASR for the singlet current \eqref{qedplusqcd} in terms of the function $B(Q^2,s_0)$ \eqref{gluon_anom_contrib1}, we get,
	
	\begin{equation}\label{mix0}
	\Sigma f_P^{(0)}F_{P\gamma}(Q^2) = \frac{N_cC^{(0)}}{2\pi^2}\frac{s_0}{s_0+Q^2} \left[1+B(Q^2, s_0) \right].
	\end{equation}
	
	Taking into account the lowest contributions, given by the $\pi^0, \eta$, and $\eta'$ mesons, the ASRs for the isovector, octet (\ref{asr38aa}), and singlet (\ref{mix0}) currents comprise a system of equations,
	\begin{equation}\label{system}
	\left(
	\begin{matrix}
	f_{\pi^0}^{(3)} & f_{\eta}^{(3)} & f_{\eta'}^{(3)} \\
	f_{\pi^0}^{(8)} & f_{\eta}^{(8)} & f_{\eta'}^{(8)} \\
	f_{\pi^0}^{(0)} & f_{\eta}^{(0)} & f_{\eta'}^{(0)} 
	\end{matrix}
	\right)
	\left(
	\begin{matrix}
	F_{\pi^0}(Q^2) \\
	F_{\eta}(Q^2)  \\
	F_{\eta'}(Q^2)
	\end{matrix}
	\right)=
	\left(
	\begin{matrix}
	\frac{N_cC^{(3)}}{2\pi^2}\frac{s_3}{s_3+Q^2} \\
	\frac{N_cC^{(8)}}{2\pi^2}\frac{s_8}{s_8+Q^2}  \\
	\frac{N_cC^{(0)}}{2\pi^2}\frac{s_0(1+B(Q^2,s_0))}{s_0+Q^2}
	\end{matrix}
	\right),
	\end{equation}
	whose solution leads to the expressions for the form factors,
	
	\begin{equation}\label{solution_generic}
	F_{P\gamma}(Q^2) = \alpha_{P} \frac{s_3}{s_3+Q^2} + \beta_{P} \frac{s_8}{s_8+Q^2} + \gamma_{P} \frac{s_0}{s_0+Q^2}[1+B(Q^2,s_0)],
	\end{equation}
	where $P = \pi^0, \eta, \eta'$.  The coefficients $\alpha_P$, $\beta_P$, $\gamma_P$ are expressed in terms of the decay constants $f_P^{(i)}$,  $\Delta$ is the determinant of the decay constant matrix in (\ref{system}):
	\begin{align}
	\small &\alpha_{\pi^0}= \frac{N_cC^{(3)}}{2\pi^2\Delta}(f^{(8)}_{\eta}f^{(0)}_{\eta'} - f^{(0)}_{\eta}f^{(8)}_{\eta'}),&\beta_{\pi^0}= \frac{N_cC^{(8)}}{2\pi^2\Delta}(f^{(0)}_{\eta}f^{(3)}_{\eta'} - f^{(3)}_{\eta}f^{(0)}_{\eta'}),&\gamma_{\pi^0}= \frac{N_cC^{(0)}}{2\pi^2\Delta}(f^{(3)}_{\eta}f^{(8)}_{\eta'} - f^{(8)}_{\eta}f^{(3)}_{\eta'}), \label{coef_pi}\\ \label{coef_eta}
	&\alpha_{\eta}= \frac{N_cC^{(3)}}{2\pi^2\Delta}(f^{(0)}_{\pi^0}f^{(8)}_{\eta'} - f^{(8)}_{\pi^0}f^{(0)}_{\eta'}),&\beta_{\eta}= \frac{N_cC^{(8)}}{2\pi^2\Delta}(f^{(3)}_{\pi^0}f^{(0)}_{\eta'} - f^{(0)}_{\pi^0}f^{(3)}_{\eta'})\;,&\gamma_{\eta}= \frac{N_cC^{(0)}}{2\pi^2\Delta}(f^{(8)}_{\pi^0}f^{(3)}_{\eta'} - f^{(3)}_{\pi^0}f^{(8)}_{\eta'}),\\ \label{coef_etaprime}
	&\alpha_{\eta'}= \frac{N_cC^{(3)}}{2\pi^2\Delta}(f^{(8)}_{\pi^0}f^{(0)}_{\eta} - f^{(0)}_{\pi^0}f^{(8)}_{\eta}),&\beta_{\eta'}= \frac{N_cC^{(8)}}{2\pi^2\Delta}(f^{(0)}_{\pi^0}f^{(3)}_{\eta} - f^{(3)}_{\pi^0}f^{(0)}_{\eta}),&\gamma_{\eta'}= \frac{N_cC^{(0)}}{2\pi^2\Delta}(f^{(3)}_{\pi^0}f^{(8)}_{\eta} - f^{(8)}_{\pi^0}f^{(3)}_{\eta}).
	\end{align}
		So, the expressions for the TFFs \eqref{solution_generic} are a consequence of the dispersive approach to axial anomaly and quark-hadron duality. In this way, it provides  theoretical grounds for the Brodsky-Lepage interpolation formula for pion TFF \cite{Brodsky:1981rp} and to some of its generalizations to the $\eta$ and $\eta'$ TFFs \cite{Feldmann:1998yc}. 
	
	The decay constants $f_P^{(i)}$ are defined as projections of the currents onto the meson states (\ref{def_f}). Since the meson states are not pure $SU(3)$ states, the corresponding decay constant matrix has nonzero off-diagonal elements. In the pseudoscalar sector, $\eta$ and $\eta'$ mesons manifest the largest mixing, which results in substantial values of $f_{\eta'}^{(8)}$, $f_{\eta}^{(0)}$ (see, e.g., \cite{Akhoury:1987ed, Ball:1995zv,Feldmann:1998vh,Escribano:2005qq,Klopot:2012hd,Escribano:2015yup}). Different mixing schemes used to determine these decay constants imply that the decay constants follow a particular symmetry (e.g., quark-flavor or octet-singlet) and effectively impose different restrictions on the decay constants \cite{Klopot:2012hd}, reducing the number of parameters which describe the mixing.  
		
	The pion dominates in the isovector current with $f_{\pi^0}^{(3)} = f_{\pi} = 0.1307 $ GeV, while its contribution to the octet and singlet currents as well as $\eta$ and $\eta'$ contributions to the isovector current are small:  $f_{\pi^0}^{(8)}, f_{\eta}^{(3)}, f_{\pi^0}^{(0)},f_{\eta'}^{(3)} \lesssim 0.01f_\pi$  \cite{Chao:1989yp, Ioffe:2007eg, Osipov:2015lva, Escribano:2020jdy}. Although in many cases the effects of $\pi^0$ mixing with the $\eta-\eta'$ system can be neglected, they are found to be relevant in the case of timelike region \cite{Khlebtsov:2016vyf}. In the present paper, we will confirm this observation.  
	
For the purposes of numerical study, we use several decay constant sets, obtained in different analyses of the mixing parameters. The constants of $\eta-\eta'$ mixing $f_{\eta, \eta'}^{(0,8)}$ evaluated in  \cite{Feldmann:1998vh,Escribano:2005qq,Klopot:2012hd,Escribano:2015yup} were supplemented with $f_{\pi^0}^{(8)}, f_{\eta}^{(3)}, f_{\pi^0}^{(0)},f_{\eta'}^{(3)}$ from the recent analysis \cite{Escribano:2020jdy}. Their values are listed in Appendix A. The corresponding values of $\alpha_{P},\beta_{P},\gamma_{P}$  for different decay constants sets are shown in Table \ref{table1}.

\begin{table}[H]
\centering
		\caption{The coefficients $\alpha_{P},\beta_{P},\gamma_{P}$ \eqref{coef_pi}, \eqref{coef_eta}, \eqref{coef_etaprime} in GeV$^{-1}$ from different analyses of the mixing parameters (see Appendix A).}
		\label{table1}
\begin{tabular}{l|c|c|c|c|c|c|c|c|c}
\hline
\multirow{2}{*}{Mix. sch.} & \multicolumn{3}{c|}{$\pi^0$} & \multicolumn{3}{c|}{$\eta$} & \multicolumn{3}{c}{$\eta'$} \\ \cline{2-10} 
			& $\alpha_{\pi^0}$ & $\beta_{\pi^0}$ & $\gamma_{\pi^0}$ & $\alpha_{\eta}$ & $\beta_{\eta}$ & $\gamma_{\eta}$ & $\alpha_{\eta'}$ & $\beta_{\eta'}$  & $\gamma_{\eta'}$    \\ \hline
  FKS98\cite{Feldmann:1998vh} & 0.274 & -0.0005 & 0.013 & -0.0015 & 0.127 & 0.144 & -0.0078 & -0.021 & 0.365 \\
  EF05\cite{Escribano:2005qq} & 0.274 & $4\times 10^{-6}$ & 0.012 & -0.0017 & 0.112 & 0.145 & -0.0071 & -0.0047 & 0.341 \\ 
  KOT12\cite{Klopot:2012hd} & 0.274 & -0.0005 & 0.014 & -0.0016 & 0.135 & 0.154 & -0.0087 & -0.021 & 0.406 \\ 
  EGMS16\cite{Escribano:2015yup} & 0.274 & -0.0004 & 0.013 & -0.0016 & 0.128 & 0.147 & -0.008 & -0.016 & 0.377 \\ 
\end{tabular}
\end{table}

	Let us investigate the properties of the function $B(Q^2,s_0)$  \eqref{gluon_anom_contrib1}, which represents the ratio strong to electromagnetic contributions, in the spacelike and the timelike regions using the experimental data for the TFFs. Our main goal is to establish the qualitative behavior of it, and as far as possible, to get its quantitative estimates too.

\section{Spacelike region }\label{sec3}
The pion TFF in the spacelike region \eqref{solution_generic} is described predominantly by the first term, while the second and the third terms can be neglected as the $\eta,\eta'$ admixtures to the isovector current are small ($\alpha_{\pi}\gg \beta_{\pi},\gamma_{\pi}$; see Table \ref{table1}). For this reason, the  $\pi^0$ TFF does not depend on $B(Q^2, s_0)$ and coincides with the expression obtained earlier \cite{Klopot:2010ke,Klopot:2012hd}. It is in fair agreement with the CELLO\cite{Behrend:1990sr}, CLEO\cite{Gronberg:1997fj}, BaBar\cite{Aubert:2009mc} and Belle\cite{Uehara:2012ag} experimental data.

Contrary to the pion TFF, $B(Q^2, s_0)$ plays an important role in the $\eta$ and $\eta'$ TFFs.  
As one can see from Table~\ref{table1}, the $\eta'$ TFF  \eqref{solution_generic} is almost completely determined by the third term ($\gamma_{\eta'}\gg \beta_{\eta'},\alpha_{\eta'}$), while the  $\eta$ TFF has a significant contribution from the second term ($\beta_{\eta}\sim \gamma_{\eta}$). Therefore, when evaluating $B(Q^2)$, to avoid additional uncertainty from $s_8$, we will rely on the $\eta'$ TFF data and check the validity of the results for the $\eta$ meson.

\subsection{Evaluation of $B(Q^2,s_0)$}

The case of two real photons $k^2=Q^2=0$ was considered earlier \cite{Khlebtsov:2018roy}, where the suppression of the strong anomaly contribution was found. In this limit the form factors are expressed in terms of the two-photon decay widths, $F_{P\gamma}(0)=\sqrt{\frac{64\pi\Gamma_{P\to2\gamma}}{e^4 m_P^3}}$.  The results for $B(0)$  evaluated from $F_{\eta'\gamma}$ for different decay constant sets are given in Table \ref{table2}. The obtained values $B(0)\ll 1$ reaffirm our previous observation: for real photons,
the strong anomaly contribution  appears to be much smaller than the electromagnetic one, as $B(Q^2)$ effectively represents their ratio.

\begin{table}[H]
\caption{The gluon to electromagnetic anomalies ratio $B(0, s_0)|_{s_0\sim 1 GeV^2}$ in the case of two real photons for different mixing schemes (analyses).}\label{table2}
\begin{tabular}{c|c|c|c|c}
   \hline
    Mixing schcheme & FKS98\cite{Feldmann:1998vh} & EF05\cite{Escribano:2005qq} & KOT12\cite{Klopot:2012hd} & EGMS16\cite{Escribano:2015yup} \\ \hline
    B(0) & 0.022 & 0.045 & -0.080 & -0.024  \\ 
\end{tabular}
\end{table}

\begin{figure}[H]
	\begin{floatrow}
		\ffigbox{\caption{The $\eta\rightarrow \gamma\gamma^*$ TFF \eqref{solution_generic} in the spacelike region for different mixing schemes with $B(Q^2)=B(0) \sim 0$ from Table \ref{table2}. The insert shows a low-$Q^2$ region.} \label{fig:7}}%
		{\includegraphics[width=0.5\textwidth]{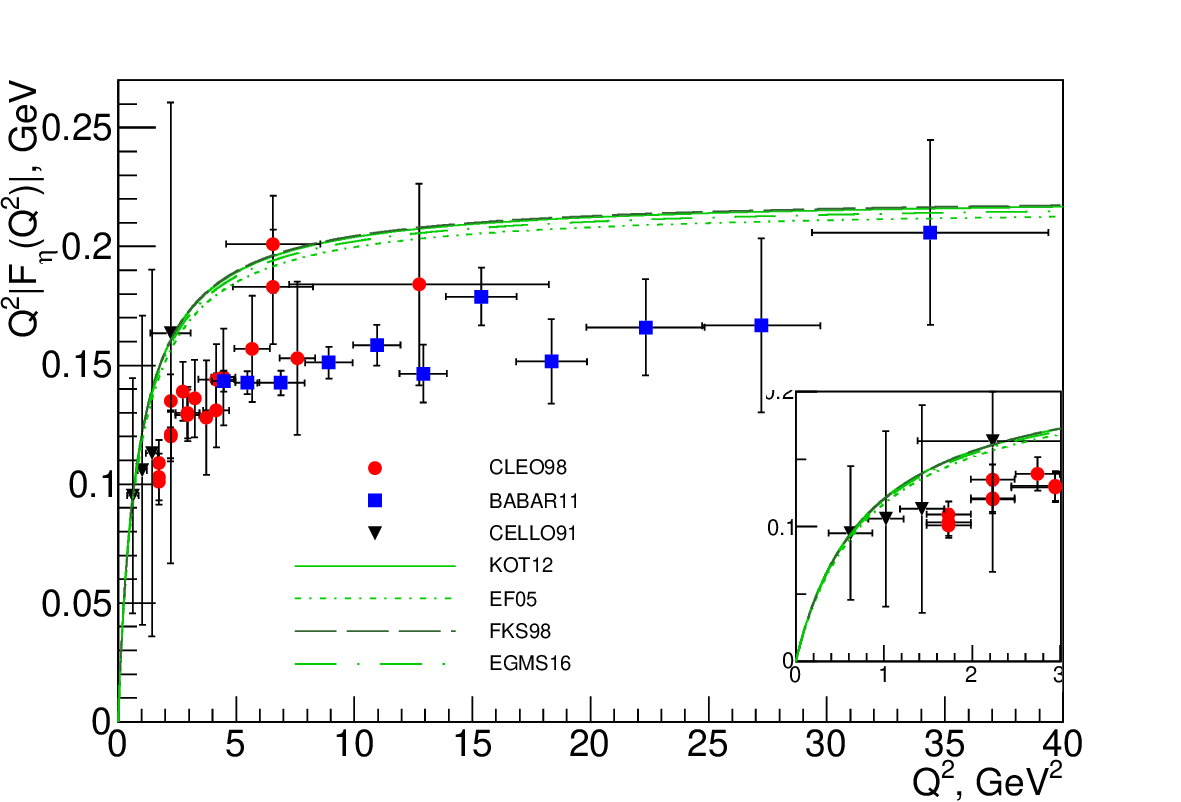}}
		\ffigbox{\caption{The $\eta'\rightarrow \gamma\gamma^*$ TFF \eqref{solution_generic} in the spacelike region for different mixing schemes with $B(Q^2)=B(0) \sim 0$ from Table \ref{table2}. The insert shows a low-$Q^2$ region.}\label{fig:8}}%
		{\includegraphics[width=0.5\textwidth]{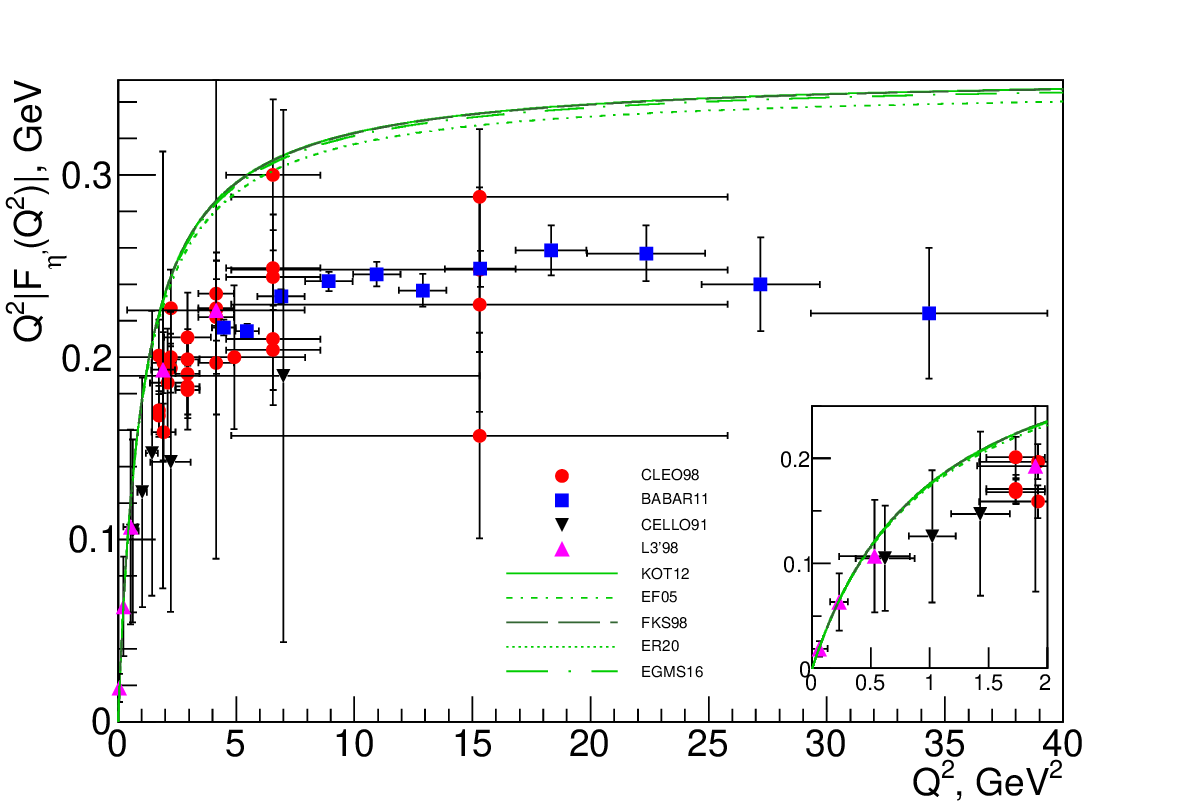}}         
	\end{floatrow}
\end{figure}
However, the assumption that the strong anomaly contribution is negligible in the case of a nonzero photon virtuality $Q^2>0$ does not hold. The plots of the $\eta, \eta'$ TFFs \eqref{solution_generic} with $B(0)$ from Table \ref{table2} (see Figs. \ref{fig:7}, \ref{fig:8}) show a strong disagreement with the experimental data, pointing out that the matrix element $\langle 0 |G\tilde{G} |\gamma\gamma^* \rangle$ should be significant. One can also observe that $Q^2|F_{\eta,\eta'}(Q^2)|$ tends to a constant value starting from $Q^2 \sim 5-10$ GeV$^2$ which indicates that $B(Q^2)$ is already close to its asymptotic value in this region.

In order to evaluate $B(Q^2)$ at $Q^2>0$, we perform a fit of the $\eta'$ TFF (\ref{solution_generic}) to the available data  from L3, CELLO, CLEO, BaBar \cite{Acciarri:1998,Behrend:1990sr,Gronberg:1997fj,BABAR:2011ad}  using the instruments of ROOT (TMinuit) \cite{James:1975dr} with  $B(Q^2)$ as a parameter (while $s_0=1$ GeV$^2$) at the intervals $Q^2\in(0.6,5.0)$~GeV$^2$, $Q^2\in(5.0,10.0)$~GeV$^2$, and $Q^2>10.0$~GeV$^2$. The obtained  values of $B$ (with the statistical errors in parenthesis and  $\chi^2_{\eta'}/dof$  in square brackets) and $B_{as}-B(0)$ for different mixing parameters are shown in Table~\ref{table3}.

\begin{table}[H]
	\caption{Results of the fit of  $\eta'$ TFF to the experimental data at different intervals of~$Q^2$.}\label{table3}
	\begin{tabular}{c|c|c|c|c}
		\hline
		Mix. sch. & $B_{(0.6-5)} [\frac{\chi^2_{\eta'}}{27-1}]$ & $B_{(5-10)}[\frac{\chi^2_{\eta'}}{9-1}]$ & $B_{as}[\frac{\chi^2_{\eta'}}{7-1}]$& $B_{as}-B(0)$  \\ \hline
		FKS98\cite{Feldmann:1998vh} & $-0.212(10)\; [0.66]$ & -0.230(8) [1.39] & -0.228(12) [0.46] & -0.250  \\ 
		EF05\cite{Escribano:2005qq}  & $-0.188(10)\; [0.65]$ & -0.206(7) [1.41]& -0.204(13) [0.46] & -0.249  \\ 
		KOT12\cite{Klopot:2012hd}  & $-0.291(10)\; [0.66]$ & -0.307(7) [1.39]& -0.306(11) [0.46] & -0.226 \\ 
		EGMS16\cite{Escribano:2015yup}  & $-0.246(10)\; [0.66]$ & -0.263(7) [1.4]& -0.262(12) [0.46] & -0.238  \\ 
	\end{tabular}
\end{table}

As one can see from Table \ref{table3}, the values of $B(Q^2)$ at $Q^2>0.6$ GeV$^2$  considerably differ from $B(0)\sim 0$ at $Q^2=0$ and are close to their asymptotical values $B_{as}$ in the region $Q^2>0.6$~GeV$^2$. The consistency of the $B$ values in different intervals supports our assumption that it is constant in this region.  Note also, that the values  $B_{as}-B(0)$ appear to be close in different mixing schemes.

\begin{figure}[H]
 \begin{floatrow}
   \ffigbox{\caption{The $\eta\rightarrow \gamma\gamma^*$ TFF \eqref{solution_generic} in the spacelike region for different mixing schemes with the approximation of $B(Q^2)$ as a step function.  The insert shows a low $Q^2$ region.} \label{fig:9}}%
           {\includegraphics[width=0.5\textwidth]{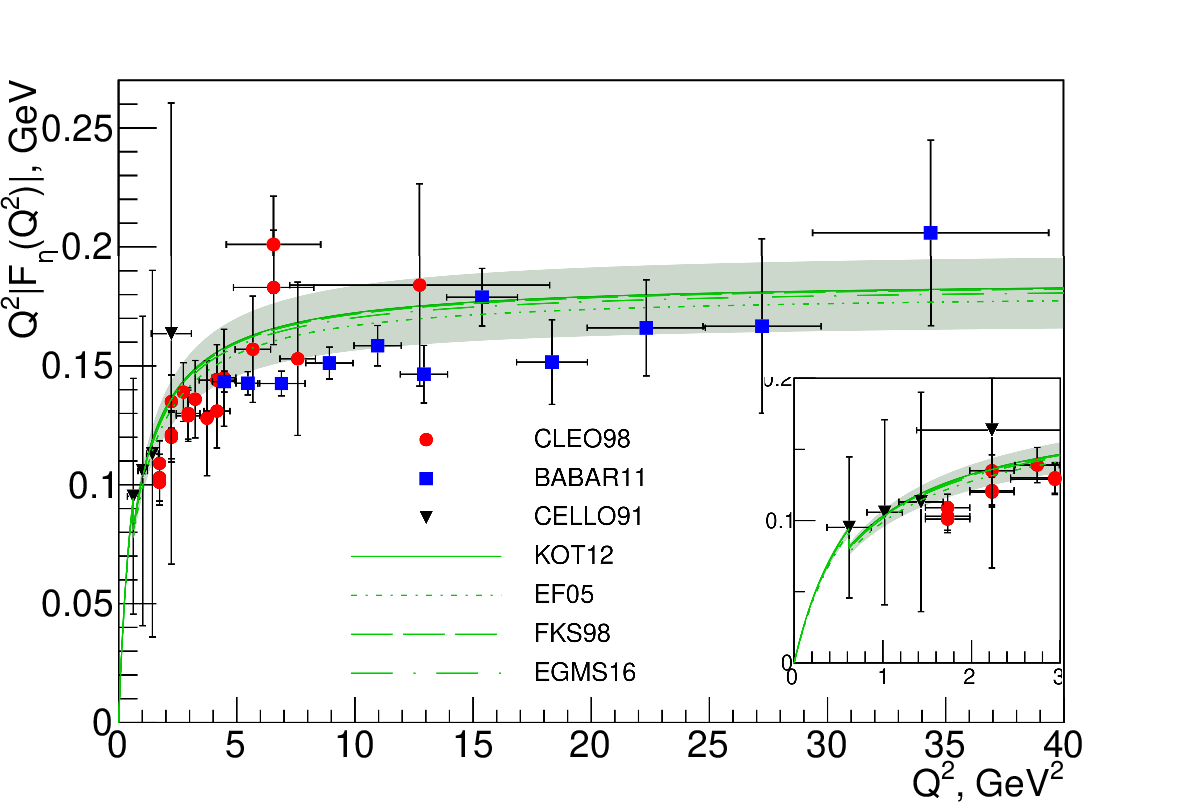}}
   \ffigbox{\caption{The $\eta'\rightarrow \gamma\gamma^*$ TFF \eqref{solution_generic} in the spacelike region for different mixing schemes with the approximation of $B(Q^2)$ as a step function.  The insert shows a low $Q^2$ region.}\label{fig:10}}%
           {\includegraphics[width=0.5\textwidth]{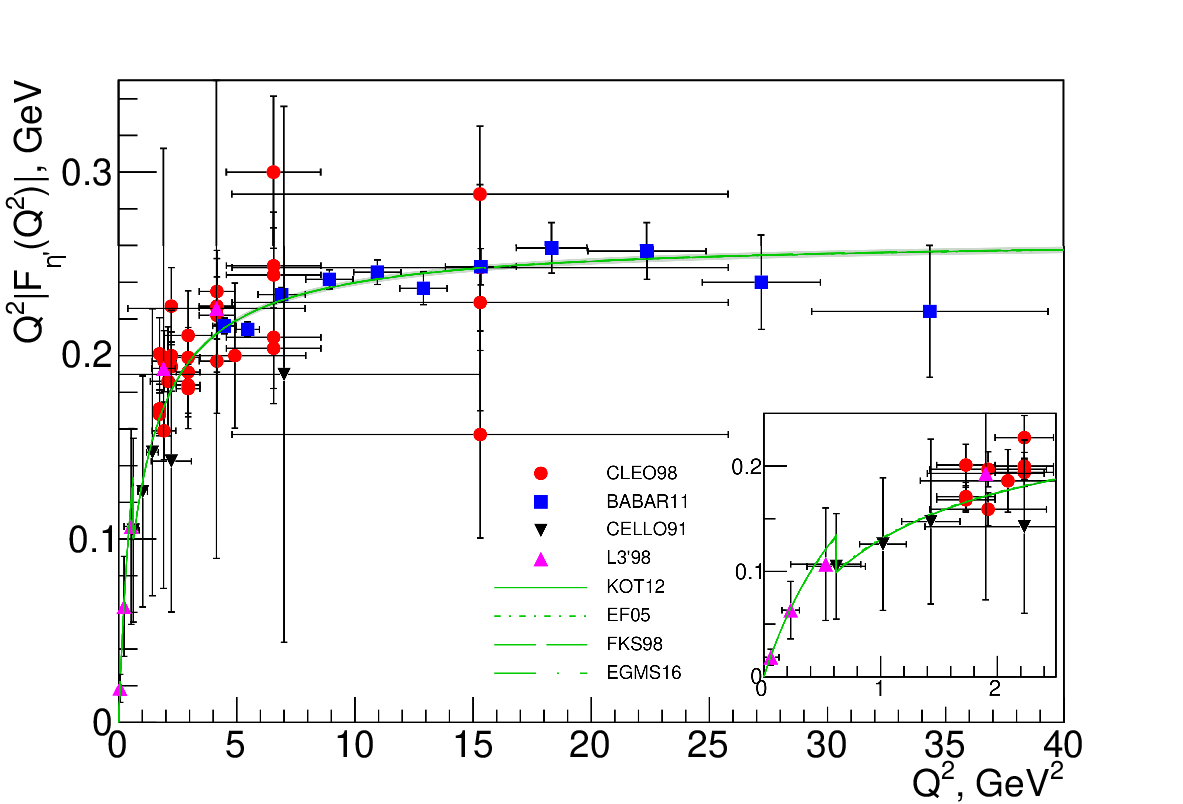}}         
  \end{floatrow}
\end{figure}

We can conclude, that at $Q^2\gtrsim 0.6$ GeV$^2$ the strong anomaly comprises $\sim 25\%$ of the electromagnetic anomaly contribution and is almost independent of the photon virtuality, while at smaller $Q^2$ it rapidly vanishes.	Taking into account that the available data at $Q^2<0.6$ are scarce and have large uncertainties, it is reasonable to test a simple approximation of the function $B(Q^2)$ by a  step-function: $B=B(0)$ at $0 \leqslant Q^2<0.6$ GeV$^2$ and $B=B_{as}$ at $Q^2\geqslant0.6$ GeV$^2$. The $\eta$ and $\eta'$ TFFs for different mixing parameters is compared with the data in Figs.~\ref{fig:9} and \ref{fig:10}. The green shaded area shows the $20\%$ uncertainty of $s_8$ (for the EGMS16 \cite{Escribano:2015yup} mixing scheme).

\subsection{$B(Q^2,s_0)$ as a function of $s_0$}
The analysis in the previous section was carried out for $s_0=1$ GeV$^2$. In order to examine the sensitivity of our approach to $s_0$, we perform a combined fit of the $\eta$ and $\eta'$ TFFs with two parameters $s_0$ and $B$ to the available data  \cite{Acciarri:1998,Behrend:1990sr,Gronberg:1997fj,BABAR:2011ad} using TMinuit \cite{James:1975dr}.  
The obtained values of $s_0$, $B$, and the corresponding $\chi^2_{\eta+\eta'}/dof$ for different mixing schemes are listed in Table \ref{table4}.
As one can see, the obtained value $s_0\simeq1$ GeV$^2$ is consistent in all considered mixing schemes.

\begin{table}[H]
\caption{$s_0$ and $B$ for different mixing schemes from two-parameter fit (with the statistical errors).
}\label{table4}
\begin{tabular}{c|c|c|c|c}
   \hline
    Mix. sch. & $s_0$ & $B$ & $B - B(0)$ & $\frac{\chi^2_{\eta+\eta'}}{dof=81-2}$  \\ \hline
    FKS98\cite{Feldmann:1998vh} & $1.00(10)$& $-0.242(62)$ & -0.264 & 1.66 \\ 
    EF05\cite{Escribano:2005qq} & $0.99(10)$ & $-0.209(66)$ & -0.254 & 1.20  \\ 
    KOT12\cite{Klopot:2012hd} & $1.01(10)$ & $-0.320(56)$ & -0.240  & 1.74 \\ 
    EGMS16\cite{Escribano:2015yup} & $1.00(10)$ & $-0.272(60)$ & -0.248  & 1.51 \\ 
\end{tabular}
\end{table}

\begin{figure}[H]
	\caption{\label{fig:11}
		Contour plot of $\chi^2_{\eta+\eta'}(B,s_0)/dof$ for $\eta, \eta'$ TFFs and the data from Refs. \cite{Acciarri:1998,Behrend:1990sr,Gronberg:1997fj,BABAR:2011ad} using the EGMS16\cite{Escribano:2015yup} mixing scheme. Dashed black contour and black filled circle denote the $99\%$ C.L. line and the minimum respectively for the combined $\eta +\eta'$ data;  the black filled square denotes $\chi^2$ minimum for the $\eta'$ data.
	}
	\includegraphics[width=0.5\textwidth]{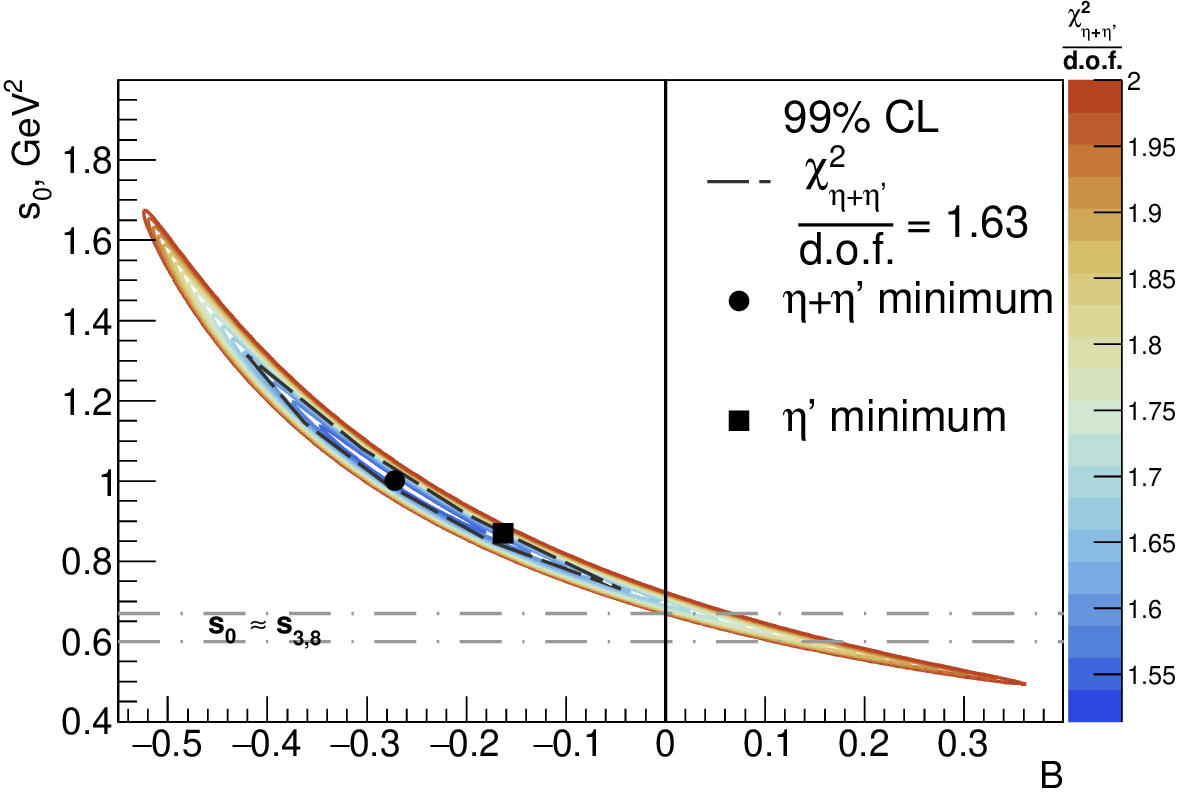}
\end{figure}

The contour plot of $\chi^2_{\eta+\eta'}/dof$ for the mixing parameters EGMS16\cite{Escribano:2015yup} is shown in Fig. \ref{fig:11}.  The results for other mixing schemes are similar. The  black filled circle and the black filled square indicate the minima of $\chi^2$ from the combined $\eta+\eta'$ and separate $\eta'$ data fits respectively; they show rather good agreement with each other.
This plot demonstrates a correlation between $B$ and $s_0$. In particular, though the best-fit values are $s_0\simeq 1$ GeV$^2$, $B\simeq-0.25$, it allows $B\sim 0$ provided $s_0 \simeq s_3\simeq 0.6$~GeV$^2$, which means  a negligible role of the strong anomaly at larger $Q^2$, not only at $Q^2=0$. We will call this scenario ($s_0\simeq 0.6$ GeV$^2$ and $B(Q^2)\simeq 0$) a \textit{hidden strong anomaly} case, as opposite to the \textit{open strong anomaly} case ($s_0\simeq 1$~GeV$^2$ and $B(Q^2)\simeq -0.25$). Further study in the timelike region will show that these regimes correspond to different kinematical regions.

The hidden anomaly scenario has a physical interpretation. The mass of the dominant contributor to the singlet ASR -- the $\eta'$ meson -- originates from the strong anomaly ($U(1)_A$ problem) \cite{tHooft:1986ooh,Diakonov:1981nv}. Neglecting the strong anomaly brings the value of $s_0$  close to the values of the octet and isovector intervals of duality $s_8$ and $s_3$, dominated by the $\eta$ and $\pi^0$ mesons.

Also, the case of hidden anomaly corresponds, in some sense, to the hypothesis \cite{Feldmann:1998vh,BABAR:2011ad} (see also \cite{Klopot:2012hd}) that an unphysical state $ (1/\sqrt{2})(|\bar{u}u\rangle+|\bar{d}d\rangle)$ differs only by a charge factor from the isovector state $ (1/\sqrt{2})(|\bar{u}u\rangle-|\bar{d}d\rangle)$.

As we already mentioned, the  singlet current is not renormalization invariant, which results in the renormalization-scale dependence of the corresponding singlet constants $f_P^{(0)}$. The model of quark-hadron duality that we apply (\ref{qhd}) leads to the decay constants fixed on the mass scales of the respective mesons in the singlet ASR. However, more refined models could result in  $Q^2$-dependence of the scale of these constants in the ASR. Numerically, $f_P^{(0)}|_{\mu^2=\infty}\simeq 0.85f_P^{(0)}|_{\mu=1 GeV^2}$ \cite{Agaev:2014wna,Escribano:2015nra,Escribano:2015yup}.  In this case, choosing  the scale as large as 49 GeV$^2$ (exceeding the currently available experimental data range), results in  $B_{as}|_{\mu^2=49GeV^2}=0.86B_{as}|_{\mu^2=1GeV^2}$ which is well within our expected theoretical uncertainty of  20~$\%$. 

\subsection{$B$ and $s_0$ from Padé approximants}
\begin{figure}[H]
	\begin{floatrow}
		\ffigbox[\FBwidth]{\caption{$s_0(Q^2)$ from the ASR \eqref{mix0} using Padé approximations for the cases of hidden (dashed green line) and open (solid red line) anomaly.} \label{pade_fig:1}}%
		{\includegraphics[width=0.5\textwidth]{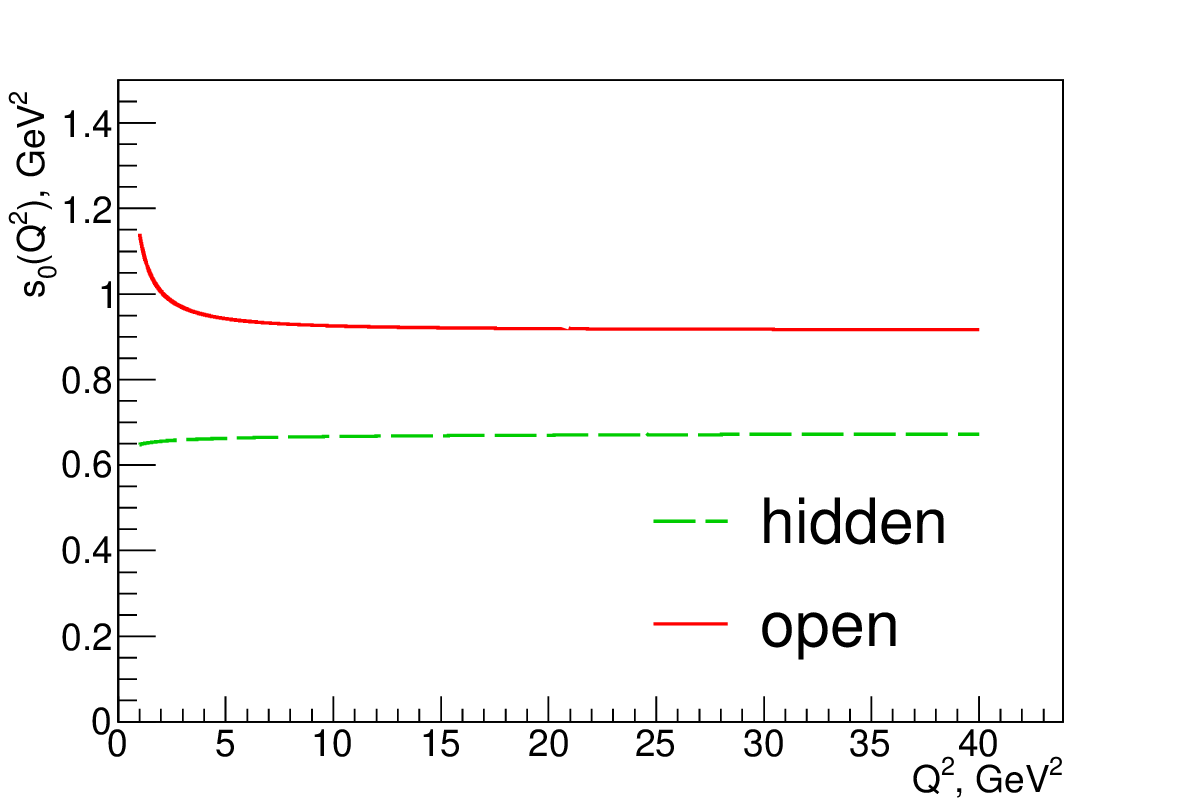}}
		\ffigbox[\FBwidth]{\caption{$B(Q^2)$ from the ASR \eqref{mix0} using Padé approximations  ($s_0=1$GeV$^2$).}\label{pade_fig:2}}%
		{\includegraphics[width=0.5\textwidth]{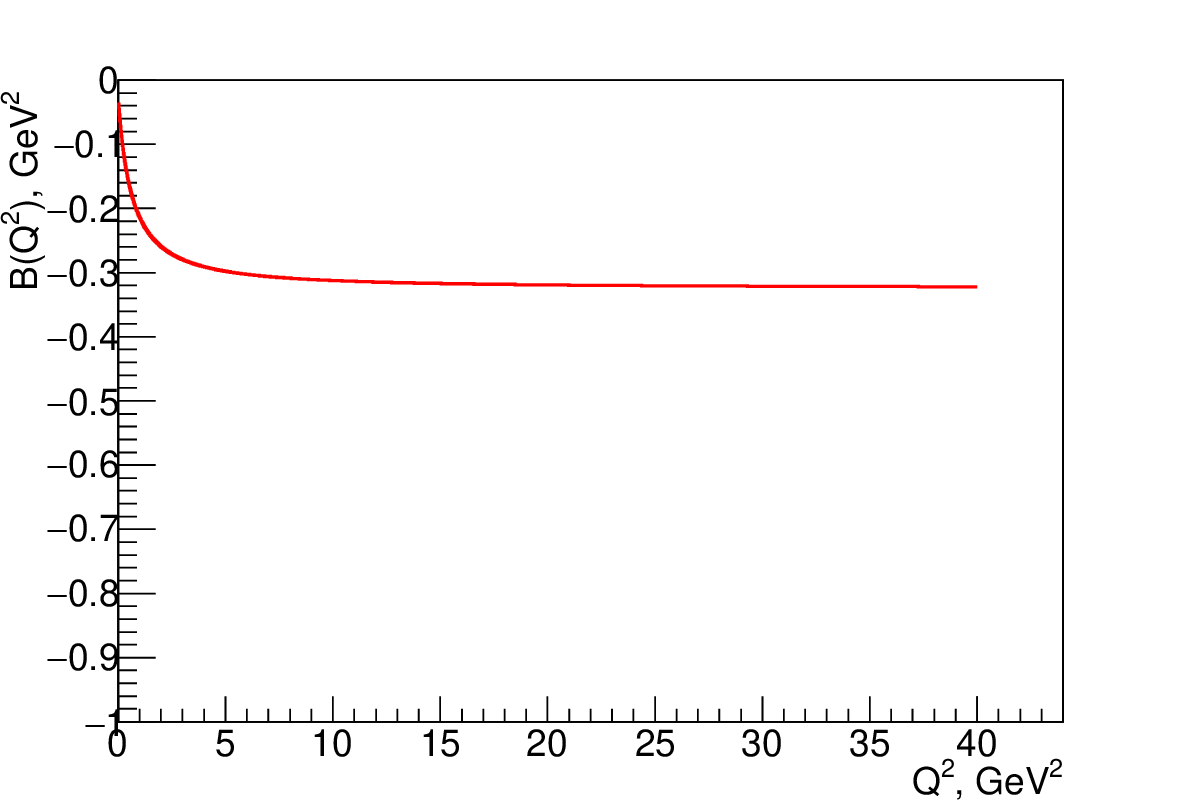}}         
	\end{floatrow}
\end{figure}

It is instructive to test our results using available approximation formulas for the TFFs. As one of such recent descriptions,  we will use the one that employs the Padé approximants \cite{Masjuan:2012wy,Escribano:2015nra,Escribano:2015yup}. In this approach, the TFFs are described by means of rational functions,
\begin{equation}\label{PL1}
	F_{P\gamma}(Q^2)=P^N_M(Q^2) \equiv  \frac{t_1 + t_2 Q^2+\cdots +t_N (Q^2)^N}{1+r_1 Q^2 + \cdots + r_M (Q^2)^M}. 
\end{equation}
The coefficients of the approximants -- $P^1_1(Q^2)$ for $\pi^0$ \cite{Masjuan:2012wy}, $\eta'$ \cite{Escribano:2015yup} and $P^2_2(Q^2)$ for $\eta$ \cite{Escribano:2015nra} -- are listed in Appendix B.  The values of  $f_{\pi^0,\eta,\eta'}^{(0)}$  were employed from the EGMS16 \cite{Escribano:2015yup} mixing scheme. 
Substituting the  TFFs (\ref{PL1}) directly into the ASR for the singlet current \eqref{mix0}, we evaluate the $s_0$ as a function of $Q^2$.
The results for $s_0(Q^2)$ for the hidden ($B=0$) and open ($B=B_{as}=-0.26$) anomaly cases are shown in Fig. \ref{pade_fig:1} by dashed green and solid red curves respectively.
One can observe that the function $s_0(Q^2)$ demonstrates close to constant behavior in both cases, confirming our evaluations:  $s_0\simeq 0.6$ is close to $s_{3,8}$  in the hidden anomaly scenario, while $s_0\simeq 1$ GeV$^2$ in the open anomaly scenario.

Also, for the open anomaly scenario ($s_0=1$~GeV$^2$), we can extract the function $B(Q^2)$  from the ASR \eqref{mix0}. The result (see Fig. \ref{pade_fig:2}) confirms our estimations: $B(Q^2)\simeq 0$ at $Q^2=0$ and rapidly reaches its asymptotic value $B\simeq -0.3$ at higher photon virtualities.

So we can see that the comparison with the Padé approximations of the TFFs \cite{Masjuan:2012wy,Escribano:2015nra,Escribano:2015yup} confirm our  conclusions made in the previous subsections.

\section{Timelike region}\label{sec4}
The analytic continuation of the the ASRs for isovector and octet currents \eqref{asr38} to the timelike region was carried out earlier \cite{Klopot:2013laa}. It was shown that for Eqs. \eqref{asr38aa}, the analytic continuation to the timelike region leads to the following relations for the real parts of the TFFs,
\begin{equation}\label{mix38}
\Sigma f_P^{(3,8)}Re F_{P\gamma}(q^2) = \frac{N_c C^{(3,8)}}{2\pi^2}\frac{s_{3,8}}{s_{3,8}-q^2}.
\end{equation} 
The corresponding imaginary parts of the TFFs are negligible (implying $|F_P(q^2)|\simeq |Re F_P(q^2)|$) everywhere except the vicinity of $q^2=s_{3,8}$ \cite{Klopot:2013laa}. 

The analytic continuation of the ASR for the singlet current \eqref{asr0} is not so straightforward. The rhs of Eq. \eqref{asr0} contains functions depending on $Q^2$ appeared due to the strong anomaly matrix element, also the spectral density $A_3^{(0)}$ contains the unknown $A_{QCD}^{(0)}$ part. 
The proper procedure of the analytic continuation of the ASR  \eqref{asr0}  is beyond the scope of this paper, however, we can try to make it directly from the equation for the TFFs \eqref{mix0}. The Eq. \eqref{mix0} differs from the analogous relations for the isovector and octet currents (\ref{asr38aa}) by the unknown function $B(Q^2)$ in the rhs In the previous chapter we established that for the spacelike photon virtualities ($Q^2=-q^2 > 0$) the function $B(Q^2)$ rapidly tends to a constant value. The asymptotic behavior of the TFFs \eqref{solution_generic} and $B(Q^2)$ for the timelike and spacelike photon virtualities should coincide. Therefore, at high $q^2$ one can suppose that the analytic continuation for the singlet case is analogous to the isovector and octet ones, and so it results in a replacement of $Q^2$ with $-q^2$,
\begin{equation}\label{mix00}
\Sigma f_P^{(0)}Re F_{P\gamma}(q^2) = \frac{N_c C^{(0)}}{2\pi^2}\frac{s_{0}[1+B(q^2,s_0)]}{s_{0}-q^2}.
\end{equation} 
We will try to describe the TFFs in the form of Eq. \eqref{mix00} at all $q^2$, not just at high $q^2$. At the same time, at intermediate region, $B(q^2)$ can be probably more complicated than in the spacelike region, e.g., it may not be monotonic. The particular behavior of $B(q^2)$ will be extracted from the data.

Solving  Eqs. \eqref{mix38} and \eqref{mix00}, we obtain the expressions similar to Eqs. \eqref{solution_generic}, but for the real parts of the TFFs in the timelike region. Therefore, with the exception of the vicinities of $q^2=s_3, s_8, s_0$, we can write for the absolute values of the timelike TFFs as
\begin{equation}\label{solution_generic_time}
|F_{P}(q^2)| = \left | \alpha_{P} \frac{s_3}{s_3-q^2} + \beta_{P} \frac{s_8}{s_8-q^2} + \gamma_{P} \frac{s_0}{s_0-q^2}\left[1+B(q^2,s_0)\right] \right |,
\end{equation}
where $P = \pi^0, \eta, \eta'$. This equation looks similar to that appearing in the vector meson dominance (VMD) model and provides a complementary justification for it based on the dispersive approach to axial anomaly (see Sec. \ref{sec-vdm}).

Because of such structure of the TFFs \eqref{solution_generic_time}, even small coefficients $\alpha_P, \beta_P, \gamma_P$ become relevant in the timelike region, in contrast to the spacelike one. This appears to be of special importance in the pion TFF: the third term with $\gamma_{\pi^0}\ll\alpha_{\pi^0}$ cannot be neglected in this case.  Moreover, as we will see, $\gamma_{\pi^0}\neq0$ is necessary for the correct description of the data, which also conforms recent analyses indicating a nonzero mixing of $\pi^0$ with the $\eta-\eta'$ system \cite{Escribano:2020jdy}. Let us note, that a similar form of the transition form factor in timelike was also employed in Ref. \cite{Khlebtsov:2016vyf}.

For the purposes of evaluation of the TFFs  (\ref{solution_generic_time}), we will use the decay constants from Refs. \cite{Escribano:2015yup,Escribano:2020jdy} (denoted as EGMS16 in Appendix A).

\subsection{Dalitz decays domain }
Let us start our study of the $\eta$ and $\eta'$ mesons in the timelike region from the Dalitz decays domain ($4m_l^2\leq q^2<m^2_P$). The $\eta$ meson experimental data are available in two decay modes: a dielectron mode, measured by A2 collaboration \cite{Adlarson:2016hpp}, and a dimuon mode, measured by NA60 collaboration \cite{Arnaldi:2016pzu}, covering the region of $0<q^2<0.25$ GeV$^2$.  Concerning  the $\eta'$ meson, the experimental data are available only in the dielectron mode from the BESIII collaboration \cite{Ablikim:2015wnx} in the region $0<q^2<0.56$~GeV$^2$.

\begin{figure}[H]
	\begin{floatrow}
		\ffigbox{\caption{The $\eta$ TFFs \eqref{solution_generic_time} for different $s_0$. The experimental data are by A2  \cite{Adlarson:2016hpp} and NA60 \cite{Arnaldi:2016pzu} collaborations.} \label{fig:14}}%
		{\includegraphics[width=0.5\textwidth]{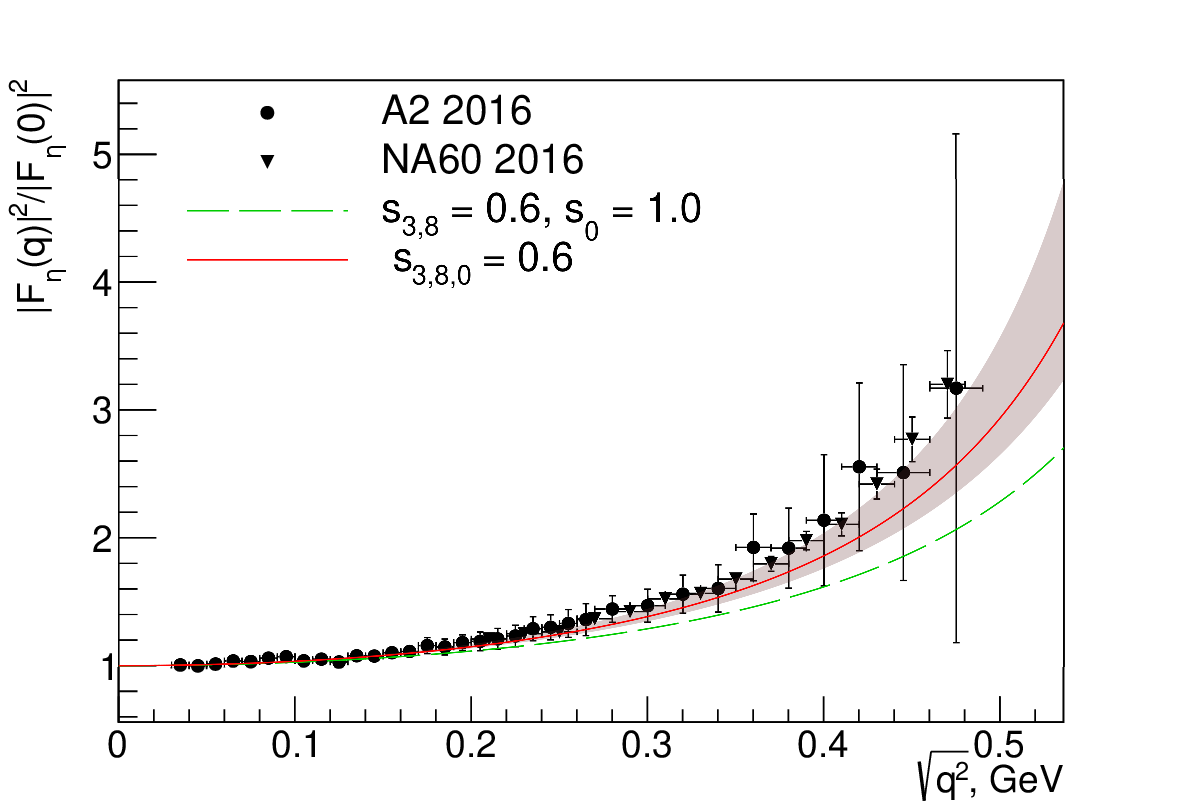}}
		\ffigbox{\caption{The $\eta'$ TFFs \eqref{solution_generic_time} for different $s_0$. The experimental data are by BESIII collaboration \cite{Ablikim:2015wnx}. }\label{fig:15}}%
		{\includegraphics[width=0.5\textwidth]{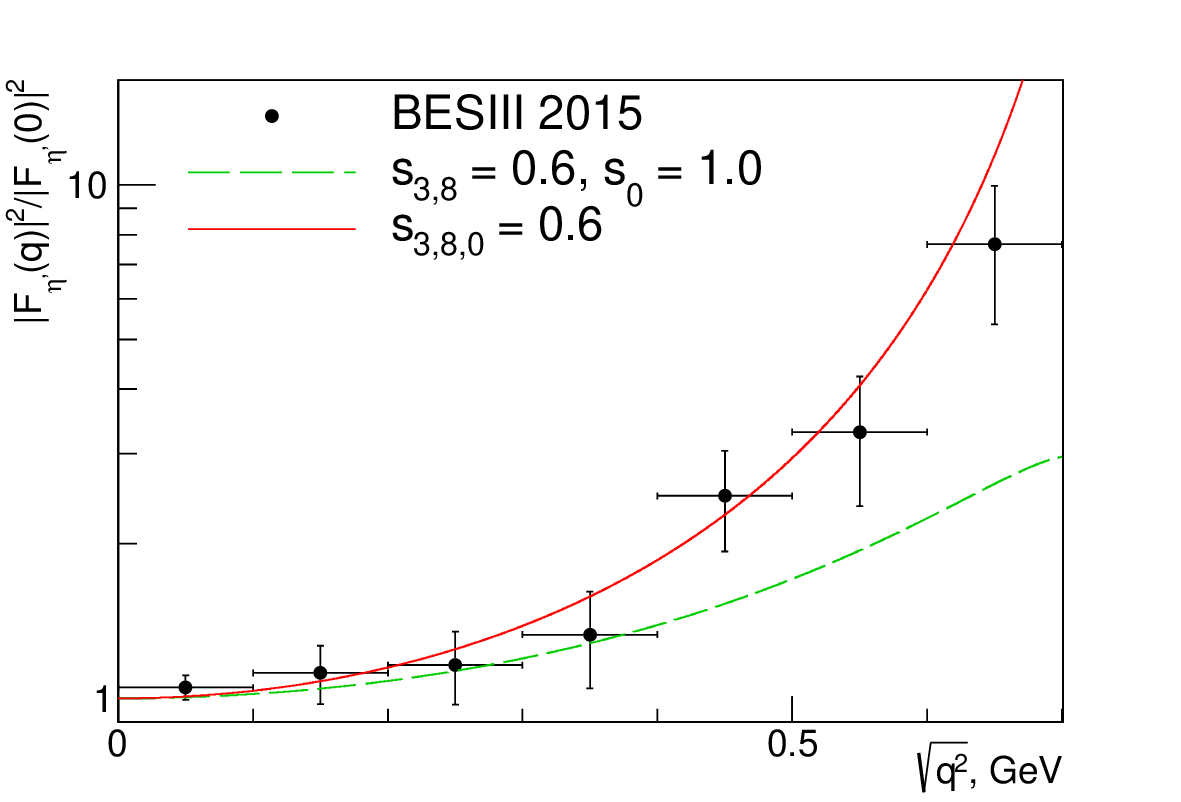}}         
	\end{floatrow}
\end{figure}

In the previous section we discussed two different cases: the hidden anomaly (negligible strong anomaly contribution and $s_0\approx s_{3,8}$) and the open anomaly (significant strong anomaly contribution and $s_0 \simeq 1$ GeV$^2$). The TFFs \eqref{solution_generic_time} for both cases (solid red and dashed green curves respectively) are shown in Figs.~\ref{fig:14} and \ref{fig:15};  the shaded area indicates the estimated theoretical uncertainty (contributed by $s_8$) of $20\%$. The respective $\chi^2$ values and the curve slopes at $q^2=0$, $a_P\equiv \lim_{q^2\rightarrow0} \frac{\partial }{\partial q^2} \frac{|F_P(q^2)|^2}{|F_P(0)|^2}$, are given in Table \ref{table5}. 
\begin{table}[H]
	\caption{$\chi^2/dof$ and curve slopes $a_P$ of the timelike $\eta, \eta'$ TFFs \eqref{solution_generic_time} for different $s_0$. 
	}\label{table5}
\begin{tabular}{l|cc|cc}
	\hline
	$s_{0}$,  GeV$^2$ & $\frac{\chi^2_{\eta}}{dof=48-1}$ & $a_{\eta}$, GeV$^{-2}$ & $\frac{\chi^2_{\eta'}}{dof=7-1}$ & $a_{\eta'}$, GeV$^{-2}$ \\ \hline
	$s_{0}=1.0$ & 5.96 & 1.4 & 1.8 & 0.95  \\		
	$s_{0}=s_{8}=0.6$ & 1.8 & 1.67 & 0.84 & 1.67  \\
	\hline
	Experiment	& & $1.97(13)$ \cite{Adlarson:2016hpp} && $1.58(34)$ \cite{Ablikim:2015wnx} \\	
		& & $1.934(84)$ \cite{Arnaldi:2016pzu}&&  \\	
	\end{tabular}
\end{table}

One can see from Figs. \ref{fig:14}, \ref{fig:15} and Table \ref{table5}, that the data in the considered  small-$q^2$ region support the hidden anomaly case. 
Let us note that this conclusion ($B=0$ at low-$q^2$) remains valid also for the  pion TFF, see the insert in Fig.\ref{fig:13}. The value of the slope $a_{\pi^0}=1.639$ GeV$^{-2}$ also agrees with the experimental value of the A2 collaboration  $1.646 \pm0.549$ GeV$^{-2}$ \cite{Adlarson:2016ykr}. 

\subsection{Annihilation domain of $\pi^0$} \label{subsec_pi_TL}

In the annihilation domain ($q^2>m^2_{\pi^0}$), the cross section of $e^+e^- \rightarrow \pi^0\gamma$ was measured with high accuracy by CMD \cite{Akhmetshin:2004gw} and SND \cite{Achasov:2016bfr} collaborations.

\begin{figure}[H]
	\caption{$\pi^0$ TFF \eqref{solution_generic_time} compared with the experimental data \cite{Adlarson:2016ykr,Achasov:2016bfr,Akhmetshin:2004gw}. The solid red curve - TFF with a Gaussian $B(q^2)$ \eqref{B_modified}, the dashed greed curve  --  TFF with a step $B(q^2)$, the dot-dashed blue curve -- mixing of $\pi^0$ and $\eta-\eta'$ is neglected. The insert shows the Dalitz decay domain.
} 
	\label{fig:13}
	\includegraphics[width=0.5\textwidth]{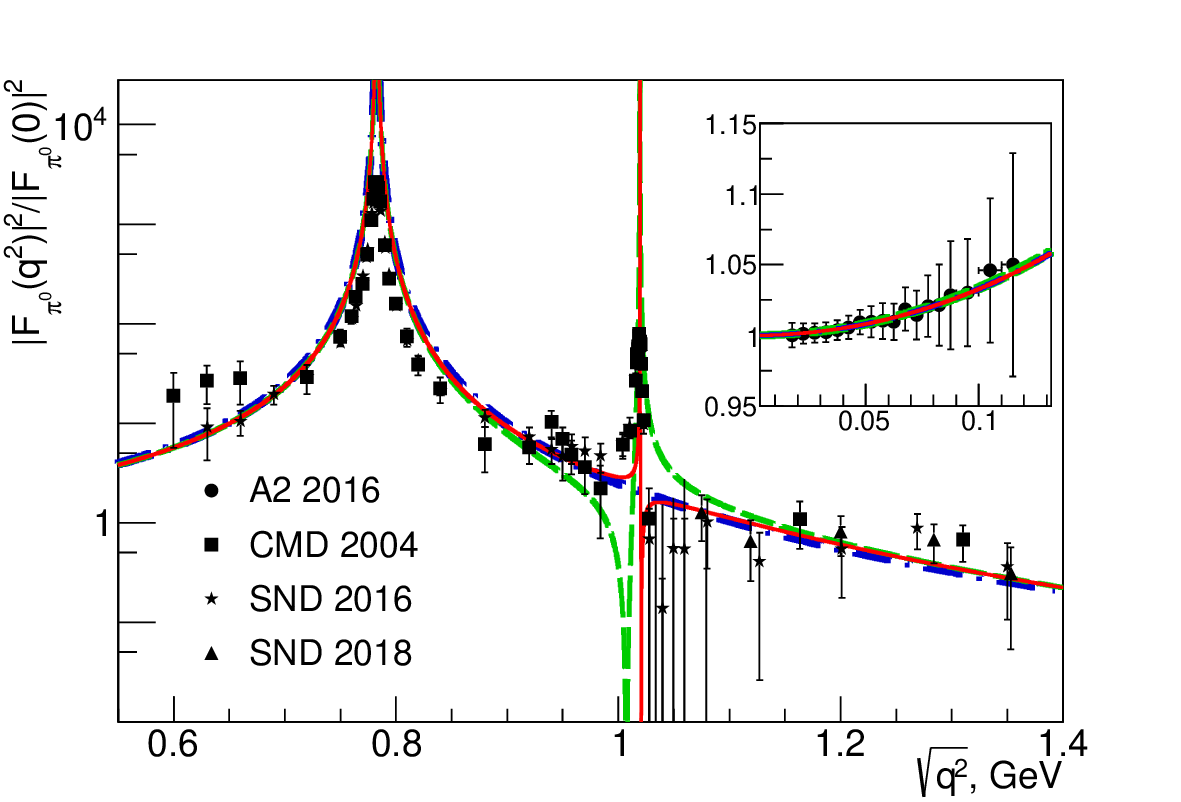}
\end{figure}

The function $B(q^2)$ at large $q^2$ in the timelike region should coincide with that in the spacelike region, $B(q^2\to \infty)=B_{as}$, while at $q^2=0$, $B(0)\simeq 0$. Let us suppose that it has the same form as the estimate obtained earlier in the spacelike region, $B(q^2)=B(0)=-0.024$ at $q^2<0.6$ GeV$^2$ and $B(q^2)=B_{as}=-0.26$ at at $q^2>0.6$ GeV$^2$ (results from Tables \ref{table2},\ref{table3} for the EGMS16 mixing scheme). Employing the same values for the duality intervals as in the spacelike region, $s_8=s_3=0.6$ GeV$^2$, $s_0=1$ GeV$^2$ at $q^2>0.6$ GeV$^2$, we plot the  corresponding $\pi^0$ TFF in Fig. \ref{fig:13} (dashed green curve) and compare it to the data \cite{Achasov:2016bfr,Akhmetshin:2004gw}. The insert shows the results for the Dalitz decay domain \cite{Adlarson:2016ykr}. 

We see that the last term in (\ref{solution_generic_time}) is necessary for reproducing the second peak in the data, implying that $\gamma_{\pi^0}$ is nonzero and even small $\pi^0$ --- $\eta$-$\eta'$ mixing has impact here. For comparison, the dot-dashed blue curve shows indicates the limit $\gamma_{\pi^0}=0$, i.e. when $\pi^0$ and $\eta$-$\eta'$ mixing is neglected.

However, the vicinity of the second peak (at $q^2\simeq 1$ GeV$^2$) is described incorrectly, so in order to improve the data description we need to adjust the function $B(q^2)$. The interference pattern of the data indicates that the first two and the third pole terms in the TFF should have different signs in this region, which can be satisfied only if $1+B(q^2)$ becomes negative at $q^2\simeq1$ GeV$^2$. Therefore, $B(q^2)$ has an extremum in this region $B<-1$. As a modification of $B(q^2)$, we suggest a Gaussian function

\begin{equation}\label{B_modified}
B(q^2) = \begin{cases} 0, & \mbox{if } 0<q^2<0.6 \ GeV^2, \\ be^{-\frac{(\sqrt{q^2} -\mu)^2}{2c^2}} + B_{as}, & \mbox{if } q^2>0.6 \ GeV^2,   \end{cases}
\end{equation}
where $B_{as} = -0.26$ is the asymptotic value, and $b, \mu, c$ are the parameters. Numerically, the parameters were found to vary, $-1.25\lesssim b <-1.00$, $0.1\lesssim c \lesssim 0.2$ GeV, $\mu \simeq 1$ GeV (we took into account the $\eta$ meson TFF constraints, which will be analyzed in the next section). The  $\pi^0$ TFF for the parameters $b=-1.1, c = 0.15$ GeV, $\mu = 1.0$ GeV is shown in Fig. \ref{fig:13} as a solid red curve. One can see that this modified $B(q^2)$ significantly improves the data description.
We also considered a different modification of $B(q^2)$, in the form of a step function. It gives a worse description, especially in the case of the $\eta$ TFF (dashed green curves in Figs. \ref{fig:13} and \ref{fig:16}). 

The suggested Gaussian modification is also more physically motivated, providing a model for nonperturbative behavior of the non-local condensates \cite{Mikhailov:1991pt}. The similar short-range behavior can be observed in the  transverse momentum dependent parton distributions \cite{Ioffe:1982tk,Teryaev:2004df}.

Therefore, we can conclude that the pion TFF experimental data in the timelike region confirms the existence of $\pi^0-\eta-\eta'$ mixing. We found that the function $B(q^2)$ should have a sharp minimum at $q^2\simeq 1$ GeV$^2$ with $B(q^2\simeq 1)<-1$.

\subsection{Annihilation domain of $\eta$ and $\eta'$ }
In the annihilation domain ($q^2>m^2_P$) the cross section of $e^+e^- \rightarrow \eta\gamma$ was measured with high accuracy by CMD \cite{Akhmetshin:2004gw} and SND \cite{Achasov:2007kw,Achasov:2013eli} collaborations for $\sqrt{s}=\sqrt{q^2} \in(0.6, 2.0)$ GeV. The processes $e^+e^- \rightarrow \eta(\eta')\gamma$ were measured also at $q^2 = 112$ GeV$^2$ by BaBar collaboration \cite{Aubert:2006cy}.

\begin{figure}[H]
	\caption{The $\eta$ TFF \eqref{solution_generic_time} compared with the experimental data \cite{Achasov:2007kw,Achasov:2013eli,Akhmetshin:2004gw}: the solid red curve -- $\eta$ TFF with Gaussian modification of $B(q^2)$ \eqref{B_modified}, dashed green curve --  without such modification.} \label{fig:16}
	\includegraphics[width=0.5\textwidth]{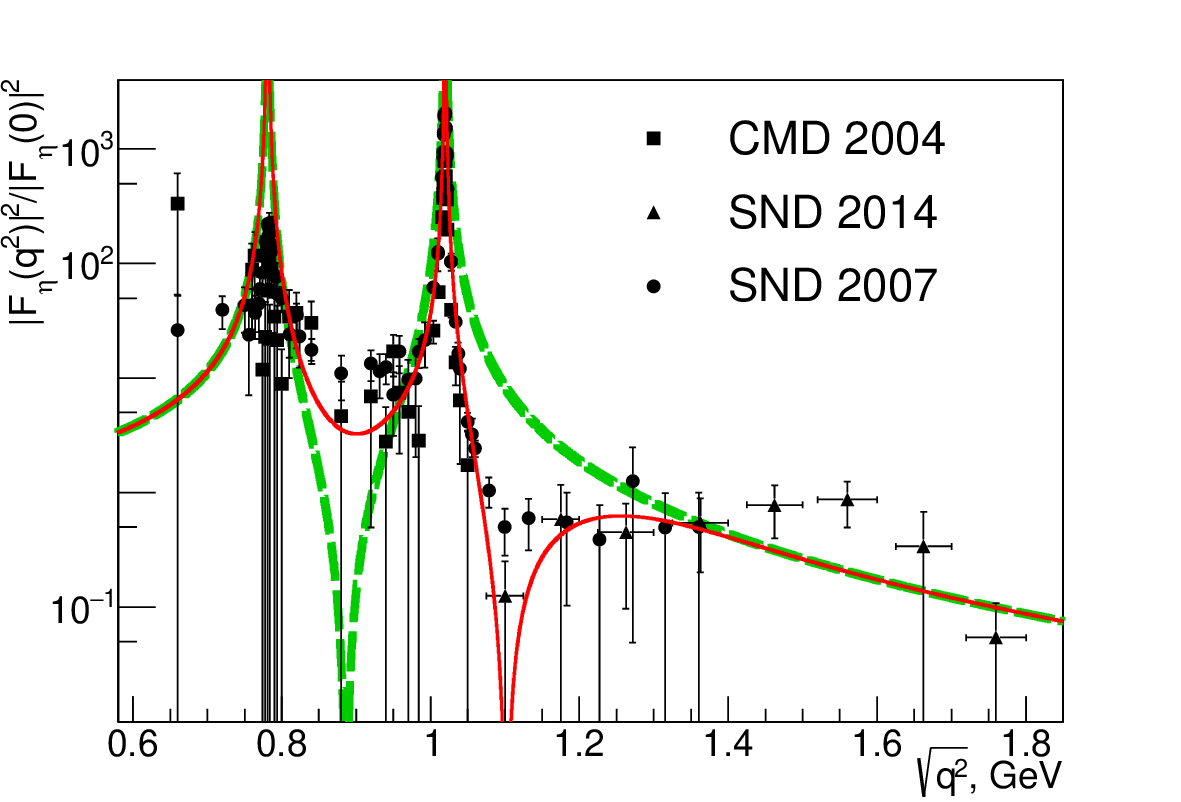}
\end{figure}

We employ the same Gaussian $B(q^2)$ discussed earlier in the pion case \eqref{B_modified}. The $\eta$ TFF with the same parameters $b=-1.1, c = 0.15$~GeV, $\mu = 1.0$~GeV (solid red curve) and the experimental data points \cite{Achasov:2007kw,Achasov:2013eli,Akhmetshin:2004gw} are shown in Fig. \ref{fig:16}. 
For comparison, the dashed green curve denotes the $\eta$ TFF without Gaussian modification of $B(q^2)$ (i.e. $b=0$).

We can conclude that the proposed $B(q^2)$ with Gaussian minimum at $q^2\simeq 1$ GeV$^2$ \eqref{B_modified} gives a consistent description of the $\pi^0, \eta$ TFFs in a wide range of $q^2$, providing correct reproduction of the data at small and large $q^2$ as well as the interference pattern near the poles.
 
The explored Gaussian-like minimum of $B(q^2)$ can mean a resonance contribution in the gluon matrix element,  $\int_{0}^{\infty} ImR(s,Q^{2})ds=\sum_X{\langle 0 |G\tilde{G} |X \rangle\langle X |\gamma\gamma^* \rangle}$. From the fact that the minimum is at  $q^2\simeq 1$ GeV$^2$ one can suppose that it is provided by pseudoscalar resonance(s) \footnote{This resonance does not lead to unobserved enhancement at low $q^2$ if its a dominating decay modes are $X\rightarrow \phi\gamma$ or $X\rightarrow \rho'\gamma$. Providing the particular example of the axial-vector duality which we discuss below \cite{Klopot:2013laa}. } with masses  $\sim 1- 1.3$ GeV, so the virtual photon energy in the final state could  be $\sim$1 GeV.  A light glueball-like state is a possible candidate. It is also possible, that this minimum is a result of a sum of several resonances.  Let us note, that some approaches \cite{Rosenzweig:1981cu,Faddeev:2003aw,Gutsche:2009jh,Cheng:2008ss,Wang:2015mla} predict the lowest glueball state with the mass $\sim1.4$ GeV, which is close to this region. However,  higher mass resonances are also often considered as candidates for the pseudoscalar glueball (see, e.g., \cite{Eshraim:2012jv, Eshraim:2019sgr}).
 
Finally, let us consider the high-$q^2$ region, where the $\eta$ and $\eta'$ TFFs were measured at the point $q^2=112$ GeV$^2$ by BaBar \cite{Aubert:2006cy}. The results for the $\eta$ and $\eta'$ TFFs \eqref{solution_generic_time} with $B=B_{as}, s_0=1$ GeV$^2$, $s_3=s_8=0.6$ GeV$^2$ at this point for different mixing schemes are listed in Table \ref{table6}. We can see a good description of the $\eta'$ data. For the case of $\eta$ meson the results are consistent with the experiment at the order of $2\sigma$.
\begin{table}[H]
\caption{$\eta$ and $\eta'$ \eqref{solution_generic_time} TFFs at $q^2 = 112$ GeV$^2$.}\label{table6}
\begin{tabular}{c|c|c}
	Mix. sch. &$q^2|F_{\eta}(q^2)|$, GeV &  $q^2|F_{\eta'}(q^2)|$, GeV  \\ \hline
    FKS98 \cite{Feldmann:1998vh} & 0.183 & 0.255 \\ 
	EF05 \cite{Escribano:2005qq} & 0.179 & 0.258 \\ 
	KOT12 \cite{Klopot:2012hd} & 0.184 & 0.256 \\ 
	EGMS16 \cite{Escribano:2015yup} & 0.183 & 0.262 \\
\hline
	Experiment \cite{Aubert:2006cy} & 0.229(31) & 0.251(20) \\ 
\end{tabular}
\end{table}

\begin{figure}[H]
	\caption{The function of the strong/electromagnetic anomaly ratio $B(q^2)$.} \label{fig:17}%
	{\includegraphics[width=0.5\textwidth]{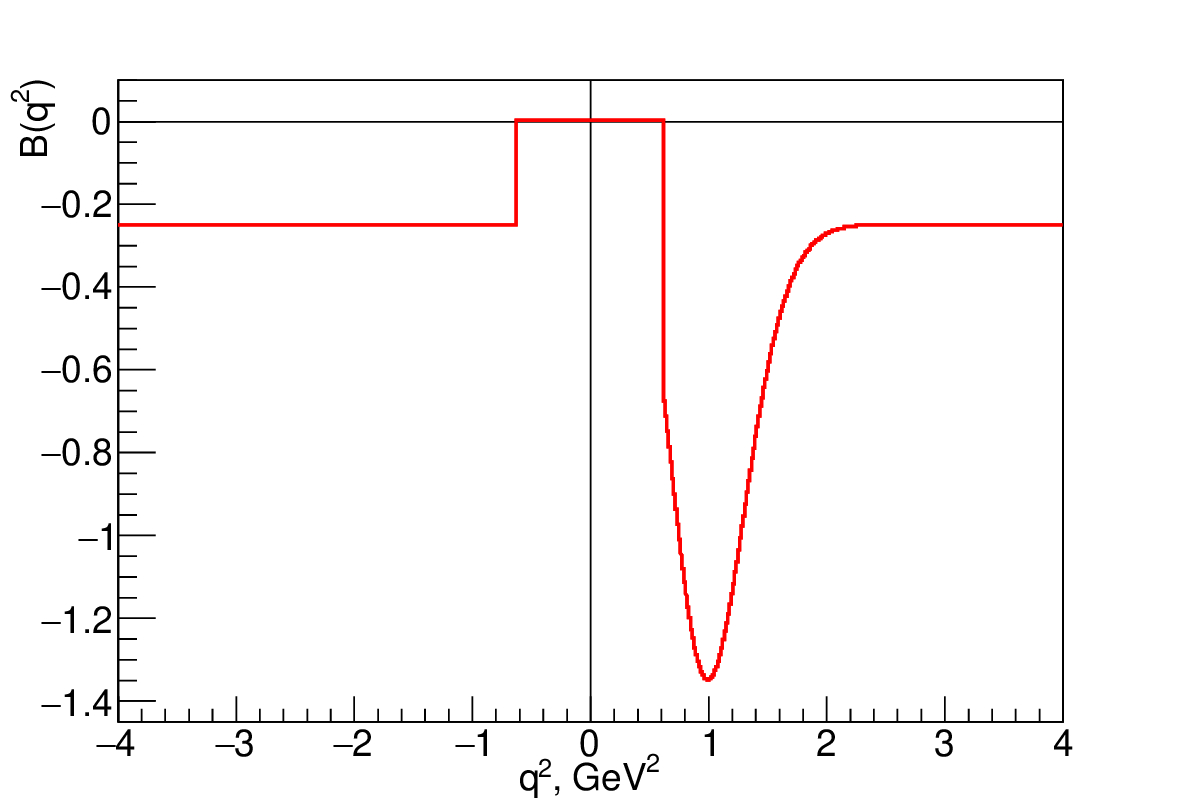}}
\end{figure}

Therefore, from the analysis of the $\pi^0$, $\eta$ and $\eta'$ meson TFFs at various photon virtualities, we can make the following conclusions for the strong to electromagnetic anomaly ratio $B(q^2)$. At low-$|q^2|$, the hidden (strong) anomaly case ($B\simeq0$ and $s_{0}\approx s_{8,3}$) takes place. At larger $|q^2|\gtrsim0.6$~GeV$^2$, the strong anomaly contribution to the TFFs rapidly becomes significant, reaching $\simeq 25\%$ of the electromagnetic one ($B\simeq-0.25$, $s_0\simeq1$~GeV$^2$). The function $B(q^2)$ has an extremum at the timelike $q^2 \simeq 1$ GeV$^2$.  Qualitatively, the function $B(q^2)$ can be described by a curve shown in Fig. \ref{fig:17}.

\subsection{Axial-vector duality \label{sec-vdm}}

As we saw in the previous section, the values of the duality intervals $s_{3,8,0}$ evaluated in the spacelike region are directly related to  the positions of the peaks in the timelike region of the TFFs.

It is worth noting that the processes $e^+e^- \rightarrow \pi^0(\eta)\gamma$ can be described within the framework of the VMD model, which implies the intermediate vector meson contributions to the photon propagator \cite{Sakurai:1960ju,OConnell:1995nse}. In such a model, a timelike pseudoscalar meson TFF is represented by a sum over the lightest vector meson contributions ($\rho, \omega, \phi$).

Comparing our results with the VMD model, we observe  correlation between the quantities inherent to the axial channel (and pseudoscalar hadron resonances) and vector hadron resonances, observed earlier  in the octet channel \cite{Klopot:2012hd, Klopot:2013laa}: $s_3, s_8, s_0\longleftrightarrow m_{\rho}^2, m_{\omega}^2, m_{\phi}^2$.

In this paper we additionally established that the mixing parameters are important for the data description in the vicinity of the peaks. 
So, the decay constants of pseudoscalar mesons (which determine $\alpha_P,\beta_P,\gamma_P$ coefficients) correspond to the vector meson coupling constants (residues) of the VMD model. This means, in particular, that the large $\eta$-$\eta'$ mixing in the pseudoscalar sector is correlated with the residues of the vector mesons.   

These observations confirm that the axial anomaly in its dispersive form (i.e. respective ASRs) reveals the duality between axial and vector channels. This duality is also related \cite{Pasechnik:2005ae} to the theorems \cite{Vainshtein:2002nv} for longitudinal and transverse parts of two-point VA correlators in external electromagnetic fields. There are also relations between resonances in different channels in the holographic approach \cite{Son:2010vc}.

\section{Conclusions}\label{last}

We considered the non-Abelian axial anomaly in the framework of dispersive approach to derive the anomaly sum rule for the singlet current \eqref{asr0}.  By means of this ASR and the ASRs for the isovector and octet currents we obtained the transition form factors of the $\pi^0,\eta,\eta'$ mesons in spacelike \eqref{solution_generic} and timelike \eqref{solution_generic_time} regions. In order to investigate the contribution of the strong (gluon) anomaly, stemming from the  matrix element $\langle 0 |G\tilde{G} |\gamma\gamma^* \rangle$, to these processes, we introduced the function $B(q^2,s_0)$, which is a ratio of the strong and  electromagnetic anomalies. This function provides the description of the nonperturbative vacuum structure and may be also studied in lattice simulations. 

We confirmed our earlier observation \cite{Khlebtsov:2018roy} that at low photon virtuality $|q^2|$ the role of the strong anomaly in the  corresponding processes is negligible ($B(q^2,s_0)$ is close to zero).  At larger $|q^2|$ in both, spacelike and timelike regions, the relative strong anomaly contribution rapidly reaches its  asymptotic value  $B\simeq -0.25$. However, in contrast to the spacelike region, in the timelike region it should have a sharp extremum (minimum $B\simeq-1.3$) at $q^2\sim1$~GeV$^2$ (Fig.\ref{fig:17}), so the strong anomaly contribution exceeds the electomagnetic one.  This might be attributed to glueball-like pseudoscalar resonances. 

We observed the correlation between the value of the duality interval for the singlet current $s_0$. The analysis showed, that when $B\simeq0$, the duality interval of the singlet current becomes close to the values of the duality intervals of the octet and isovector currents, $s_0\simeq s_3\simeq s_8 \simeq 0.6$ GeV$^2$.	
This (hidden anomaly) scenario occurs in the region of small $|q^2|$, while at larger $|q^2|$ the relative strong anomaly contribution $B$ becomes significant while $s_0$ reaches $1$ GeV$^2$ (open anomaly).

From our analysis we found that the mixing of the pion with the $\eta$ and $\eta'$ mesons is necessary for correct description of the pion timelike TFF. This is in agreement with recent estimations of mixing parameters from $VP\gamma$ decays, where a nonzero isospin-symmetry breaking was evaluated \cite{Escribano:2020jdy}. 

The axial anomaly in its dispersive form reveals the duality between axial (pseudoscalar) and vector sectors: the intervals of duality of the axial channel were found to be numerically close to the vector meson masses, while the combinations of the mixing parameters and the intervals of duality correspond to the behavior near the poles and the coupling constants (residues) of the VMD model.
 
\textbf{Acknowledgements. }We are indebted to  Andrey Kotov, Maria Paola Lombardo, Sergey Mikhailov and Aleksander Osipov for useful discussions and comments. The paper was supported by Russian Science Foundation Grant No. 21-12-00237.
\section*{Appendix A}\label{appendix_a}
We list here  the values of decay constants used in the present work and which were evaluated in several other analyses: the decay constants of the $\eta-\eta'$ mixing ($f^{(8,0)}_{\eta,\eta'}$) were taken from Refs. \cite{Feldmann:1998vh,Escribano:2005qq,Klopot:2012hd,Escribano:2015yup}, and the constants of the $\pi^0$ admixtures to the $\eta-\eta'$ system ($f^{(3)}_{\eta,\eta'}$ and $f_{\pi}^{(8,0)}$) were taken from \cite{Escribano:2020jdy}. The mixing parameters evaluated in the cited works were expressed in terms of the decay constants.  The pion decay constant is $f_{\pi^0}^{(3)}=f_{\pi}=0.1307$~GeV.

\begin{equation} \label{fks98}
	\textbf{FKS98}\cite{Feldmann:1998vh}:\left(
	\begin{matrix}
		f_{\pi^0}^{(3)} & f_{\eta}^{(3)} & f_{\eta'}^{(3)} \\
		f_{\pi^0}^{(8)} & f_{\eta}^{(8)} & f_{\eta'}^{(8)} \\
		f_{\pi^0}^{(0)} & f_{\eta}^{(0)} & f_{\eta'}^{(0)} 
	\end{matrix}
	\right)=\left(
	\begin{matrix}
		1 & -0.0015 & -0.035 \\
		-0.0066 & 1.17 & -0.46 \\
		0.034 & 0.19 & 1.15 
	\end{matrix}
	\right) f_{\pi},
\end{equation}

\begin{equation} \label{EF05}
	\textbf{EF05}\cite{Escribano:2005qq}:\left(
	\begin{matrix}
		f_{\pi^0}^{(3)} & f_{\eta}^{(3)} & f_{\eta'}^{(3)} \\
		f_{\pi^0}^{(8)} & f_{\eta}^{(8)} & f_{\eta'}^{(8)} \\
		f_{\pi^0}^{(0)} & f_{\eta}^{(0)} & f_{\eta'}^{(0)} 
	\end{matrix}
	\right)=\left(
	\begin{matrix}
		1 & -0.0015 & -0.035 \\
		-0.0066 & 1.39 & -0.59 \\
		0.034 & 0.054 & 1.29 
	\end{matrix}
	\right) f_{\pi},
\end{equation}

\begin{equation} \label{KOT12}
	\textbf{KOT12}\cite{Klopot:2012hd}:\left(
	\begin{matrix}
		f_{\pi^0}^{(3)} & f_{\eta}^{(3)} & f_{\eta'}^{(3)} \\
		f_{\pi^0}^{(8)} & f_{\eta}^{(8)} & f_{\eta'}^{(8)} \\
		f_{\pi^0}^{(0)} & f_{\eta}^{(0)} & f_{\eta'}^{(0)} 
	\end{matrix}
	\right)=\left(
	\begin{matrix}
		1 & -0.0015 & -0.035 \\
		-0.0066 & 1.11 & -0.42 \\
		0.034 & 0.16 & 1.04 
	\end{matrix}
	\right) f_{\pi},
\end{equation}

\begin{equation} \label{EGMS16}
	\textbf{EGMS16}\cite{Escribano:2015yup}:\left(
	\begin{matrix}
		f_{\pi^0}^{(3)} & f_{\eta}^{(3)} & f_{\eta'}^{(3)} \\
		f_{\pi^0}^{(8)} & f_{\eta}^{(8)} & f_{\eta'}^{(8)} \\
		f_{\pi^0}^{(0)} & f_{\eta}^{(0)} & f_{\eta'}^{(0)} 
	\end{matrix}
	\right)=\left(
	\begin{matrix}
		1 & -0.0015 & -0.035 \\
		-0.0066 & 1.18 & -0.46 \\
		0.034 & 0.14 & 1.13 
	\end{matrix}
	\right) f_{\pi},
\end{equation}

\begin{equation} \label{EGMS16_renorm}
	\textbf{EGMS16 renormalized at 49 GeV$^2$}\cite{Escribano:2015yup}:\left(
	\begin{matrix}
		f_{\pi^0}^{(3)} & f_{\eta}^{(3)} & f_{\eta'}^{(3)} \\
		f_{\pi^0}^{(8)} & f_{\eta}^{(8)} & f_{\eta'}^{(8)} \\
		f_{\pi^0}^{(0)} & f_{\eta}^{(0)} & f_{\eta'}^{(0)} 
	\end{matrix}
	\right)=\left(
	\begin{matrix}
		1 & -0.0015 & -0.035 \\
		-0.0066 & 1.18 & -0.46 \\
		0.032 & 0.13 & 1.06 
	\end{matrix}
	\right) f_{\pi},
\end{equation}
\section*{Appendix B}\label{appendix_b}
The Padé approximants' coefficients for $\pi^0$, $\eta$, and $\eta'$ mesons,  their masses, and two-photon decay widths:

$\pi^0$: $t_1 = 0.277, \ r_1 = 1.778\; (P^1_1(Q^2))$ \cite{Masjuan:2012wy}; $m = 0.13498$~GeV, $\Gamma_{ \pi^{0}\to \gamma \gamma} = 7.8\cdot10^{-9}$~GeV~\cite{Zyla:2020zbs}.

$\eta$: $t_1 = 0.274, \ t_2 = 0.0469, \ r_1 = 2.0893, \ r_2 = 0.2373 \; (P^2_2(Q^2))$ \cite{Escribano:2015nra}; $m = 0.54786$~GeV,$\Gamma_{\eta\to\gamma \gamma} = 0.516\cdot10^{-6}$~GeV~\cite{Zyla:2020zbs}.

$\eta'$: $t_1 = 0.3437, \ r_1 = 1.4412\; (P^1_1(Q^2))$ \cite{Escribano:2015yup}; $m = 0.95778$~GeV,  $\Gamma_{ \eta'\to\gamma \gamma} = 4.354\cdot10^{-6}$~GeV~\cite{Zyla:2020zbs}.

\end{document}